\documentclass[]{raa}            
\usepackage{graphicx,times}
\usepackage{natbib}

\providecommand{\bm}[1]{\mbox{\boldmath$#1$\unboldmath}}

\begin{document}

   \title{Low-ionization galaxies and evolution in a pilot survey up to z = 1
    \footnotetext{Based on observations obtained in service mode at the European 
         Southern Observatory at Paranal}}
         
          \volnopage{ {\bf 2010} Vol.\ {\bf 10} No. {\bf XX}, 000--000}
         \setcounter{page}{1}

   \author{E.~Giraud \inst{1} \and  
           Q.-S.~Gu \inst{2} \and 
           J.~Melnick \inst{3} \and
           H.~Quintana \inst{4} \and
           F.~Selman\inst{3} \and 
           I.~Toledo\inst{4} \and 
           P.~Zelaya \inst{4} }
           
           \institute{LPTA, Universit\'e Montpellier 2 - CNRS/IN2P3, 34095 Montpellier, France 
	   \and Department of Astronomy, Nanjing University,
                Nanjing 210093, P. R. China 
         \and  European Southern Observatory, Alonso de C\'ordova 3107,
                Santiago, Chile 
          \and Department of Astronomy and Astrophysics, P. Universidad 
                Catolica de Chile, Casilla 306, Santiago, Chile \\
           \vs \no
           {\small Received [year] [month] [day]; accepted [year] [month] [day] }
             }

\abstract{
We present galaxy spectroscopic data on a pencil beam of  $10.75' \times7.5'$ centered on the X-ray cluster RXJ0054.0-2823 at 
$z = 0.29$. We study the spectral evolution of galaxies from $z=1$ down to the cluster redshift in a magnitude-limited sample at $\rm R\leq23$, for which the statistical properties of the sample are well understood. We divide emission-line galaxies in star-forming galaxies, LINERs, and Seyferts by using emission-line ratios of [OII], $\rm H\beta$, and [OIII], and derive stellar fractions from population synthesis models. We focus our analysis on absorption and low-ionization galaxies. For absorption-line galaxies we recover the well known result that these galaxies have had no detectable evolution since $z\sim0.6-0.7$, but we also find that in the range $z=0.65-1$ at least 50\% of the stars in bright absorption systems are younger than 2.5Gyr. Faint absorption-line galaxies in the cluster at $z = 0.29$ also had significant star formation during the previous 2-3Gyr, while their brighter counterparts seem to be composed only of old stars.  
At $z\sim0.8$,  our dynamically young cluster  had a truncated red-sequence. This result seems to be consistent with a scenario where the final assembly of E/S0 took place at $z<1$. 
In the volume-limited range $0.35\leq z\leq0.65$ we find that 23\% of the early-type galaxies have LINER-like spectra with $\rm H\beta$ in absorption and a significant component of A stars. The vast majority of LINERs in our sample have significant populations of young and intermediate-aged stars and are thus not related to AGN, but to the population of `retired galaxies' recently identified by Cid-Fernandes et al. (2010) in the SDSS.  Early-type LINERs with various fractions of A stars, and E+A galaxies appear to play an important role in the formation of the red sequence. 
\keywords{cosmology: observations -- galaxies: evolution - large scale structures - evolution
-- RX J0054.0-2823 } 
}

   \authorrunning{E. Giraud et al. }            
   \titlerunning{Low-ionization galaxies at intermediate $z$ }  
   \maketitle


\section{Introduction}

In the course of an investigation of the diffuse intergalactic light in X-ray emitting clusters at intermediate redshifts 
\citep{Melnick:1999p1185}, we detected a puzzling S-shaped arc-like structure in the ROSAT cluster RX J0054.0-2823 
\citep{Faure:2007p4999},  which we tentatively identified  as the gravitationally lensed image of a 
background galaxy at a redshift between z=0.5 and z=1.0.  The cluster, however, is characterized by  having three dominant D or cD 
galaxies in the center, two of which are clearly interacting. We designed an observing strategy that allowed us at the same time to observe 
the arc, the diffuse Intra-Cluster Light (ICL), and a magnitude limited sample of individual galaxies in the field taking advantage of the multi-object 
spectroscopic mode of the FORS2 instrument on Paranal.  By optimizing the mask design (see below) we were able to obtain: 
(a) very deep observations of the arc;
(b) very deep long-slit observations of the ICL; and
(c) redshifts and flux distributions for 654 galaxies of which 550 are in the pencil beam and at $0.275 \leq z \leq 1.05$.

Our pencil beam sample covers a redshift range up to z = 1 (with some galaxies up to z = 1.7).
In standard cosmology with $H_o = 75$~$\rm km~s^{-1}~Mpc^{-1}$, $\Omega_{0,m}=0.30$, and $\Omega_{0,\Lambda} = 0.70$,
this range provides a large leverage of about 3000~Mpc or 7~Gyr, which should  be sufficient to extract
some of the most conspicuous characteristics on galaxy evolution at $z \leq 1$. 
About half of all stars seem to be still forming, mostly in disks, in this redshift range \citep{Dickinson:2003p4993, Hammer:2005p5042}.
Our spectroscopy provides a 50-60\% complete sample of  the galaxies in a pencil beam of $\sim 10'\times10'$,
centered on the cluster, uniformly down to R=23. Our sample compares in size with the DEEP1 spectroscopic pilot survey \citep{Weiner:2005}
but is smaller than large surveys such as DEEP2 (e.q. Lin et al. 2008; Yan et al. 2009), VVDS (e.q. Franzetti et al. 2007; Garilli et al. 2008), GOODS
(e.q. Bell et al. 2005; Weiner et al. 2006). The advantage of a pilot survey is that
it can be handled rather easily by a single (or a few) researcher(s) to test new methods, new ideas before
applying these new methods to large samples.

The vast majority of our individual spectra reduced to zero redshift have S/N ratios per $\rm 2.6 \AA$ pixel larger than 3
at $\rm 4200 \AA$. This resolution is very well adapted to the detection of small equivalent width [OII] emission, which is expected to be found
in bulge dominated galaxies with small disks, in some LINERs, in ``mixed'' mergers between E/S0 and star-forming objects,
and perhaps in some post-starbursts galaxies. The line of sight of our field crosses three main structures: a dynamically young
cluster at $z = 0.29$, an over-dense region with layers at $z = 0.4-0.5$, and a mixed region of field and possible layers from $z = 0.6$
to $z = 1$. According to morphology-density relations \citep{Dressler:1980p4383, Dressler:1997p4998, Melnick:1977p5232, Smith:2005p4379, 
Postman:2005p5116, Cooper:2006p4926, Scoville:2007p5138}, we expect that over-dense regions will provide a rather large number of red objects 
available to our study. Therefore red objects with or without star formation, or with low photo-ionization is the subject which we will focus on, having
in mind the possible roles of E+A galaxies \citep[][ and references therein]{Dressler:1983, Norton:2001, Blake:2004, Goto:2007, Yang:2008} 
and of LINERs \citep{Yan:2006} in the building-up of the red sequence.

We focus on galaxies with either low star-formation
or low ionization which appear at $z \leq 0.6$. We use line ratio diagnosis based upon [OII], $\rm H\beta$, 
and [OIII], from Yan et al. (2006), to classify galaxies in LINERs, star-forming galaxies, and Seyferts. 
This method, combined with visible morphology, allow us to isolate a significant population
of early-type LINERs, and galaxies with diluted star-formation in later morphological
types at $z = 0.35 - 0.6$.

Several studies suggest that the bulk of 
stars in  early-type cluster galaxies had a formation redshift of $z \geq 3$, 
while those in lower density environments may have formed later, but still at 
$z \geq 1.5-2$ \citep[for reviews see][]{Renzini:2006p5117, Renzini:2007p5120}. 
This may be in contradiction with the rise in the number of massive red galaxies
found by Faber et al. (2007) who concluded that most early types galaxies reached
their final form below $z =1$. Our data include a clear red sequence at  $z = 0.29$ 
and a quite large number of absorption systems up to $z \sim 1$ which we fit
with population synthesis models in order to search for age variations with
$z$ and luminosity.

The paper is structured as follows. Section~2 presents details of the 
observations and the data reduction procedures. Section~3 is on the resulting 
redshift catalog. Section~4 presents an overview of variations in spectral 
energy distribution with redshift for absorption and emission systems. Section~5 is dedicated to population variations with $z$ and luminosity
in absorption systems. Low-ionization galaxies are
in \ref{liners}. In Section \ref{redlimit} we suggest a scenario in which early-type LINERs will
become E/S0 galaxies once the A stars die, and photo-ionization disappear. Summary
and Conclusions are in Section~6.


\section{Observations and data reduction}

\begin{figure}
\centering
   \vspace*{.0cm}
   \includegraphics[width=11cm]{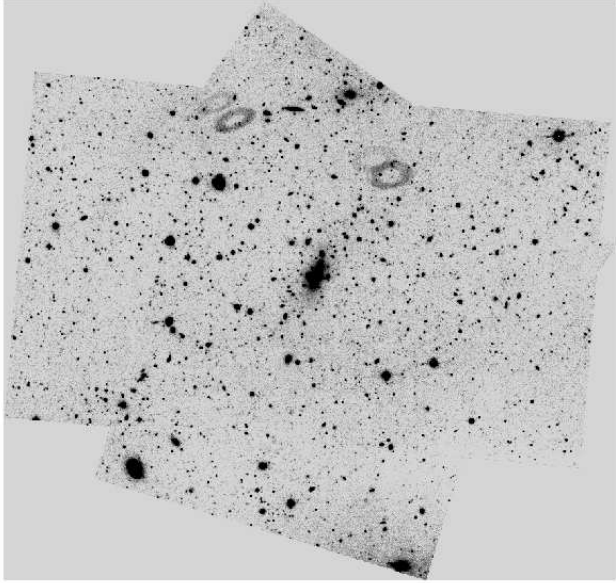}
   \vspace*{-0.0cm}
    \caption{The central (pencil beam) field from R images obtained with the wide field camera at the 2.2m telescope in La Silla}
    \label{field}
\end{figure}

The observations (ESO program 078A-0456(A) were obtained with the FORS2 instrument \citep{fors:2005} on the Cassegrain focus
of the VLT UT1 telescope in multi-object spectroscopy mode with the exchangeable mask unit (MXU).
They were acquired  in service observing and were spread over two periods 
78 and 80 to satisfy our observing conditions. FORS2  was equipped with
two $\rm 2k \times 4k$ MIT CCDs  with $15\rm \mu m$ pixels. These CCDs have high efficiency in
the red combined with very low fringe amplitudes. We used the grisms 300V and  600RI, both with
the order sorting filter GG435. With this filter, the 300V grism has a central wavelength 
at 5950~\AA\ and covers a wavelength between $4450 - 8700$~\AA\ at a resolution 
of 112~\AA$~{\rm mm}^{-1}$. The 600RI grism has a central 
wavelength of 6780~\AA\ and covers the 5120-8450~\AA\ region at a 
resolution of 55~\AA~${\rm mm}^{-1}$. Combined with a detector used in binned 
mode, the 300V grism has a pixel resolution of 3.36~\AA~pixel$^{-1}$.
The grisms were used with a slit width of 1$''$. In order to match the major and minor axis of the ICL
and the prominent arc-like feature rotation angles of $-343^o$, $-85^o$,
and $-55^o$ were applied. The slit lenghts used for the ICL spectra are  56.5$''$, 32.5$''$, 
and 24.5$''$, while those of typical galaxies vary between 7$''$ and 12$''$.
 The ICL was located either on the master CCD or the second one, resulting
in a combined pencil beam field of $\rm 10.75' \times 7.5'$ (Figure~\ref{field}).

A total of 30 hours of observing time including field acquisition, mask positioning, and integration
time were dedicated to our pencil beam redshift survey of the J0054.0-2823 field.  Each mask was filled
with 39-49 slitlets in addition to the ICL slits. In order to trace some of the apparent structures
connected to J0054.0-2823, and to reach beyond its Virial radius, we also obtained MXU exposures of
8 FORS2 fields of $\rm 7' \times 5'$ adjacent to the pencil beam,
so in total we obtained spectra of 730 individual sources.


\subsection{Mask preparation}

Tables for preparing the masks and instrument setups were obtained with the FORS Instrumental Mask Simulator\footnote{
http://www.eso.org/sci/observing/phase2/FORS/FIMS.html} \citep{FIMS:2006}.
The selection of the objects for the preparation of the slit masks of the pencil beam field was done by using a photometric
catalog in V and I which we had derived from deep images obtained in a previous NTT run \citep{Faure:2007p4999},
and pre-images in R from the VLT. The selection of the objects in the fields adjacent to the pencil beam were obtained
by using  images taken with the WIde Field Imager (WIFI; $34' \times 33'$) at the 2.2m telescope on La Silla.
Photometry in V and R from the WFI images are used throughout the paper. 
The allocated time was divided in observing blocks (OBs) to be executed in service mode.
A typical OB of 1h execution time had a science integration time of 2900s in two exposures of 1450s. 

We estimated exposure times for E to Sb galaxies in the range z = 0.3 - 0.8. 
Using the exposure time calculator of FORS, we obtained  magnitude limits,
the major steps of which are given in  Table~ \ref{maglimit}, which we
used to optimize the distribution of slitlets in the masks. 
After isolating bright objects which did not require long exposure time,
we prepared a grid with an exposure time step $\rm 2 \times 1450~s$ which
we filled with galaxies having V magnitudes 
such that the expected S/N ratio would be better than 2.8 (1 pixel along
the dispersion). After receiving VLT pre-images in the red band, we did 
a similar grid in R and adjusted
the two grids. The masks were prepared interactively with the FIMS tool 
and the R pre-images. We started to fill masks with objects that require 
an exposure time $\rm \leq 2 \times 1450~s$, then moved to $\rm \leq 4,~ 6, 
~and ~8 \times 1450~ s$. Because we prepared sets of masks with slits in 
very different directions (those of the ICL long and short axis in particular),
 objects that could not be targeted with a mask in a given direction (i.e. 
such as any mask with running name ICL-s in Table~ \ref{journal}) were 
targeted in a perpendicular one (i.e. masks with running name ICL-L), an 
approach which made the mask filling quite efficient, in particular in 
over-dense areas and field edges. Objects which were close to a predicted 
S/N of 2.8 in an OB, were selected to be also observed in another OB as
often as possible. Some objects with expected good S/N in an OB, were 
re-observed in another OB when there was no other target in the corresponding 
slit strip. They provides a set of high S/N $(\sim 20)$ ratio spectra. \rm

A total of 973 slitlets were selected, 621 in 14 different masks in the pencil beam
field, and 352 in 8 masks in the  adjacent fields. Thirty five percent of the sources
of the pencil beam field were observed through different masks, whereas the slitlets
of the adjacent fields are all for different sources.


\begin{table*}
\renewcommand{\arraystretch}{0.9}
\centering 
\caption{ Table used for preparing MXU plates of multiple Observing Blocks }
\begin{tabular}{ c c c c }
Number of OBs of 1h 	&  Integration time & Magnitude limit in V & S/N for S0-Sb at $0.3 \leq z \leq 0.8$ \\ \hline
1                   &    2900s          & 24.4 - 24.8         & 2.8 - 5.2                     \\
2                   &    5800s          & 24.8 - 25.2         & 2.8 - 5.2                     \\
4                   &   11600s         & 25.2 - 25.6         & 2.8 - 5.2                      \\
\hline \hline
\label{maglimit}
\end{tabular}
\end{table*}

The resulting list of masks and OBs, and the journal of  observations are given in
Table~ \ref{journal}. Spectra of the pencil beam  field were obtained through masks with
running names Bright, ICL-L, ICL-s,  and arc. ICL-L and ICL-s were obtained with rotator
angle $-343^o$ and $-85^o$ respectively, and arc with a rotation of $-55^o$.
Masks with names  SE, E, NE, N, NW, W, SW1 \& SW2 are on adjacent fields.
The observations were obtained during clear nights, with seeing between $0.7''$ and $1.5''$ and dark sky.


\begin{table*}
\renewcommand{\arraystretch}{0.9}
\centering
\caption{\label{journal} Journal of the MXU Observations}

\begin{tabular}{ l l l c c l}

Name & OB ID & Date & Exp. time (s) & \# slitlets & Grism\\
\hline

Bright1 & 255728 & 20 Oct. 06    & $3 \times 550$ & 34 & 600RI\\
Bright2 & 255726 & 23 Oct. 06    & $3 \times 550$ & 38 & 600RI\\
SW1 &  255710 &   18 Oct. 06    & $3 \times 550$ & 45  &  300V\\
SW2 &  255708 &  15 Oct. 06     & $3 \times 550$ & 40 &  300V \\
W   &   255712 &  19 Oct. 06     & $3 \times 550$ & 49 &  300V \\
SE & 255706 & 3 Oct. 07   & $3 \times 550$ & 42 &  300V\\
N & 255716 & 5 Oct. 07   & $3 \times 550$ & 42 &  300V\\
NW & 255714 & 5 Oct. 07   & $3 \times 550$ & 48  &  300V\\
NE & 255718 & 14 Oct. 07   & $3 \times 550$ & 47 &  300V \\
E   &  255704 & 15 Oct. 06 &  $3 \times 710$ & 39  & 600RI\\
ICL-s1 & 255750 & 12 Dec. 06  &  $2 \times 1450$ & 46&  300V \\
ICL-s2 & 255748 & 15 Nov. 06  &  $2 \times 1450$  & 48 &  300V\\
ICL-L1 &  255761 &  12 Dec. 07  &  $2 \times 1450$ & 41 &  300V\\
ICL-L2 & 255763 &  9 Jan. 07 &  $2 \times 1450$ & 39 &  300V\\
arc2 & 255734 &  24 Nov. 06  &  $2 \times 1450$ & 48 &  300V\\
arc1 & 255736, 38 & 9 Jan. 07, 11 Sept. 07 &  $4 \times 1450$ & 43&  300V \\
ICL-s3 & 255744, 46, 47 &  27 Oct. 06, 9 Nov. 06  & $6 \times 1450$ & 47&  300V \\
ICL-L3 & 255752, 59, 60 & 21 Sept. 07, 31 Oct. 07  &  $6 \times 1450$ & 49&  300V \\
arc3 & 255730, 32, 33 &  17 Aug. 07  &  $6 \times 1450$ & 46 &  300V \\
ICL-s4 & 255739, 41, 42, 43 &  23 Oct. 06, 13 Nov. 06   & $8 \times 1450$ & 46&  300V \\
ICL-L4 & 255754, 56, 57, 58 &  15, 17 \& 20 Nov. 06  & $8 \times 1450$ & 49 &  300V\\
RI &  255720, 22, 23, 24, 25 &  13 Nov. 06, 12 \& 14 Sept. 07,  &  &  &  \\
   &                         &   \& 3 Oct. 07 & $10 \times 1450$ & 47 & 600RI \\
\hline \hline
\end{tabular}
\end{table*}

\subsection{Spectral extraction}

The data were reduced by the ESO quality control group who provided us with science products
(i.e. sky subtracted, flat fielded and wavelength calibrated spectra of our objects), together
with calibration data:  master bias (bias and dark levels, read-out noise), master screen
flats (high spatial frequency flat, slit function), wavelength calibration spectra from He-Ar
lamps, and a set of spectrophotometric standards, which were routinely observed. The sky subtracted
and wavelength calibrated 2D spectra allowed a very efficient extraction of about 60 \% of the spectra.
Nevertheless the pipeline lost a significant fraction of objects, in particular when they were located
on the edges of the slitlets. To increase the efficiency of the spectral extraction we performed a new
reduction starting from frames that were dark subtracted, flat-fielded and wavelength calibrated,
but not sky subtracted, using a list of commands taken from the LONG context of the MIDAS package. 
For each slitlet, the position of the object spectrum was estimated by averaging 500 columns in the
dispersion direction between the brightest sky lines and measuring the maximum on the resulting profile.
The sky background was estimated on one side of the object, or on both, depending on each case.
Spatial distortion with respect to the columns was measured on the sky line at 5577~\AA\ and used
to build a 2D sky which was subtracted to the 2D spectrum. Multiple exposures where then aligned and
median averaged. The 1D spectra of objects were extracted from 2D medians by using the optimal
extraction  method in MIDAS.

\subsection{Redshift identification}

\begin{figure}
\centering
   \vspace*{-0.0cm}
   \includegraphics[width=11cm]{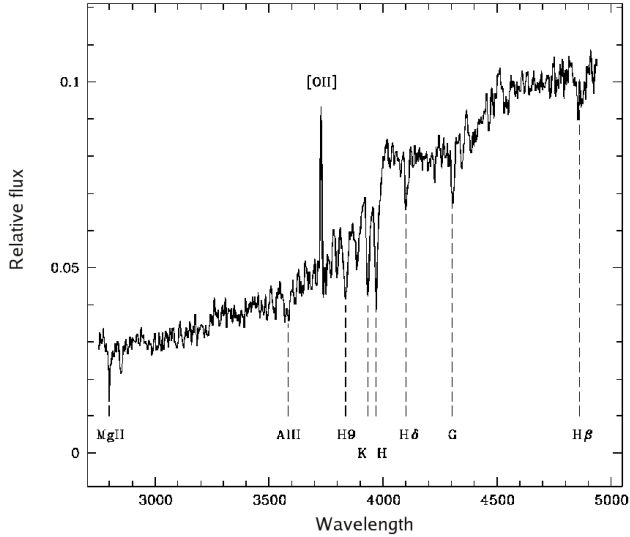}
   \vspace*{-0.0cm}
    \caption{Example of a spectrum of a red and bright galaxy with [OII] and
the main absorption lines identified}
    \label{typicalspectrum}
\end{figure}

The identification of lines for determining the redshifts was done independently by two methods
and three of the authors. The 2D spectrum was visually scanned to search for a break in the continuum,
or an emission-line  candidate (e.g. [OII]~$\lambda$3728.2~\AA). A plot of the 1D spectrum was displayed 
in the corresponding wavelength region to search for [OII],  the Ca H \& K lines, and/or Balmer
lines H$\epsilon$, H9~$\lambda$3835.4~\AA,  H8~$\lambda$3889.1~\AA,
H10~$\lambda$3797.9~\AA, and H$\delta$. The redshift was then confirmed by
searching for the [OIII] doublet $\lambda$4958.9 \&\ 5006.8~\AA, and  H$\beta$
in emission if [OII] had been detected, or G and the Mgb band, if the 4000~\AA\ break
and (or) the H and K lines had been identified. The MgII~$\lambda$2799~\AA\  line in absorption
and, in some cases AlII~$\lambda$~3584~\AA, were searched to confirm a potential redshift $z \geq 0.65 $,
while in the cases of low redshift candidates we searched for $\rm H \beta$, the NaD doublet
$\lambda$5890~\&~5896~\AA, and in a few cases H$\alpha$. The H$\gamma$\ line,
the E (FeI+CaI~$\lambda$5270~\AA) absorption feature and, in some bright galaxies
the Fe~$\lambda$4383~\AA, Ca~$\lambda$4455~\AA, Fe~$\lambda$4531~\AA\ absorption lines,
were used to improve the redshift value. The resulting identification ratio of galaxy redshifts is of the order of 90\%.
The 10\% of so-called unidentified include stars, objects with absorption lines which were not understood,
a few objects with low signal, and defects. Six QSO's were also found. 
An example of good spectrum of red galaxy, with its main absorption lines 
identified, is shown in Figure~\ref{typicalspectrum}.


A second independent visual identification was performed using Starlink's Spectral Analysis Tool (SPLAT-VO),
matching an SDSS reference table of emission and absorption lines\footnote{http://www.sdss.org/dr5/algorithms/linestable.html}
to the spectra. After a first estimate of the redshift a cross-correlation was performed using the FXCOR task on the RV package
of IRAF\footnote{IRAF is distributed by the National Optical Astronomy
 Observatory, which is operated by the Association of Universities
 for Research in Astronomy, Inc., under cooperative agreement with
 the National Science Foundation.}. Due to the large span of redshifts, two sets of
templates were used. The first one consisting of 3 template spectra of galaxies ($\lambda=3500-9000$\AA\ with emission and absorption
lines and a dispersion of 3\AA/pix) from the SDSS survey\footnote{http://www.sdss.org/dr2/algorithms/spectemplates/index.html} with
continuum subtraction using a spline3 order 5 fitting function. The second set of templates were two average composite spectra of early
type and intermediate type galaxies ($\lambda=2000-7000$\AA\  with only absorption lines and a dispersion of 2\AA/pix) from the
K20 survey\footnote{http://www.arcetri.astro.it/$\sim$k20/spe\_release\_dec04/index.html} using a spline3 order 7 function for
continuum subtraction. An interactive selection of the wavelength range used in the cross correlation was done on each spectrum
avoiding contamination by sky lines. The spectra were re-binned to the template dispersion (smaller for 300V spectra and larger for
600RI spectra) , which gave the best results. 
Velocity errors were determined from the quality of the cross-correlation, 
by using standard R value of Tonry \& Davies \citep{Tonry:1979p5170}. Here 
we used $R_T$ to differentiate it from the R band magnitudes symbol. These 
values are provided in the IRAF task FXCOR and explained in the reference 
quoted. In brief, $R_T$ is proportional to the ratio of the fitted peak height 
and the antisymmetric noise as defined by Tonry \& Davies (1979). The redshifts, 
$R_T$ values, and velocity errors are given in Table 5, which also includes 
the list of visually identified lines.

A third independent visual inspection was carried out  when a discrepancy was seen between the previous two sets of measurements,
and also in the very few cases were no redshift could be measured. For these spectra we first tried to detect emission or absorption
lines and then used Gaussian fits to establish the line centroids and their errors and shifts. The redshift of each line was measured
independently and the galaxy redshift was obtained from the weighted average of all lines. This third inspection resolved nearly all
the few remaining discrepancies so we have retained the cross-correlation values  whenever possible. We note that Xcorr failed in two
instances: 1)  for $z > 0.8$ galaxies with low S/N and few weak absorption lines, and, 2) when no absorption lines, but 1, 2 or 3 clear
emission lines were present. In these cases we used the visual line identifications and    assigned a conservative error of 300 km/s.

\begin{figure}
\centering
   \vspace{-0.0cm}
   \includegraphics[width=11cm]{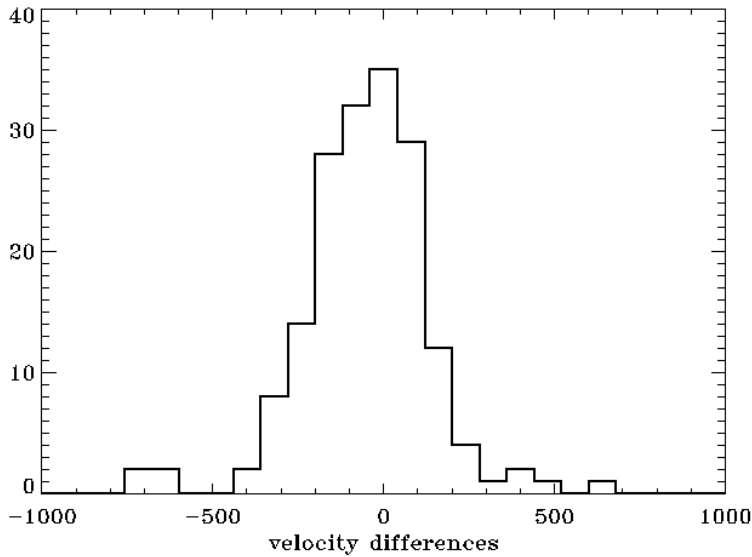}
   \vspace*{-0.0cm}
    \caption{ Radial velocity differences for galaxies with multiple observations. The objects with velocity discrepancies larger than $400$~km/s are broad line QSO's and one high-z galaxy.}
    \label{multiple}
\end{figure}

\begin{figure}
\centering
   \vspace*{-0.0cm}
   \includegraphics[width=11cm]{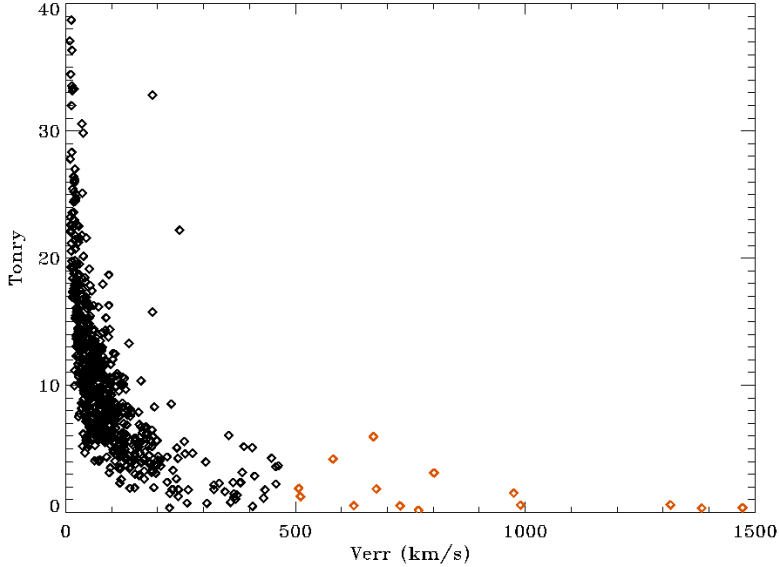}
   \vspace*{-0.0cm}
    \caption{Relation between radial velocity errors  ($V_{err}$) and the Tonry $R_T$ parameter \citep{Tonry:1979p5170} in redshifts obtained by cross-correlation.
    The points away from the general trend (5 points with $R_T>10$ and 5 with $V_{err }> 500~ \rm km~s^-1$) are 5 QSO's and distant weak spectrum galaxies with emission lines. Objects with $V_{err }> 500~ \rm km~s^-1$, marked in red, 
were not used in combined spectra. }
    \label{Tonry}
\end{figure}

Spectra from more than one mask were obtained for 94 objects. Their final velocities and velocity errors were calculated as error-weighted means
from multiple observations, although  no significant disagreements were found. These repeated observations serve as a check on the
internal errors. Figure~\ref{multiple} presents the differences between the cross-correlation velocity measurements for all galaxies
with multiple observations. The representative full width half maximum (FWHM) error is 200 km/s. In Figure~\ref{Tonry} we have plotted the
relation between velocity errors and the  Tonry $R_T$ value obtained in our cross correlations.
Most errors are $< 300$ km/s even for $4 > R_T > 2$ and the typical error is of order 80 km/s
with the vast majority of the radial velocities have errors below 200~km/s.
We have only discarded a few values with $R_T<1$ when there were no measurable emission lines.
 

 \subsection{Flux Calibrations}

The 1D spectra were divided by the response curve of the detector, which had been determined from 4 spectrophotometric standard 
stars observed along the runs, and reduced by the same method (bias, flat field, wavelength calibration, and extraction) as the 
galaxy spectra. The thick absorption telluric band of O$_2$ 
centered at 7621~\AA\ (unresolved line series) was not removed from the observation response curve 
and was considered as a feature of the global wavelength dependent efficiency. 

The relative fluxes per wavelength of the corrected spectra can be compared with stellar population  models,
in arbitrary unit, but are not calibrated in flux. The spectra were re-binned to the z = 0 rest frame with
relative flux conservation.  Because a significant fraction of spectra have a too low S/N ratio for a meaningful
comparison with population synthesis models, one may either select the brightest objects or combine spectra of
similar types. The spectra taken at  different locations of the MXU masks have different lengths along the
dispersion direction. In order to merge them the spectra were normalized to have the same flux in the region
4050-4250~\AA\ (see below).

\subsection{Quality of the spectra}

The final S/N ratio of the extracted spectra, corrected for the response 
curve, 
and re-binned to zero redshift depends on a number of parameters: seeing, night sky transparency and background, magnitude 
of the object and integration time, wavelength of the S/N measurement, and redshift. To give an idea of the final
products we present in  Table~\ref{quality} a representative set of 28 spectra at various z, magnitudes, number of OBs and resulting S/N ratio
measured on zero redshift spectra in the wavelength range 4150-4250~\AA\ which corresponds well to the location
where we will measure the main indexes of this work. S/N ratios of spectra
re-binned to zero redshift are for a pixel element of $\rm 2.6~\AA$ throughout
the paper. Table~\ref{quality} gives also the names of the OBs.


\begin{table*}
\renewcommand{\arraystretch}{0.9}
\centering
\caption{\label{quality} Signal-to-noise ratio of representative spectra. The columns indicate respectively: the redshift $(z)$ of a selected object, its V and R Petrosian magnitudes, the number of observing blocks, N(OB), from which its spectrum is extracted,  the S/N ratio measured in the wavelength range 4150-4250\AA\  of the spectrum rebinned to zero redshift, the name of observing blocks from Table~\ref{journal}, and the grism used. Spectra from OB's with running name ``arc'' have on the average higher S/N ratio than those with name ``ICL'' as illustrated by the two objects marked (*).}

\begin{tabular}{ c c c c c l l}
$z$      &   V  &  R      &  N(OB) & S/N      & Name of OBs  & Grism \\ \hline
0.2923 & 19.2 & 18.6    & 2             & 14       & arc1           		& 300V \\
0.2932 & 20.3 & 19.3    & 2             & 18       & ICL-L1 \& L2   	& 300V \\
0.2928 & 21.6 & 20.5    & 3             & 17       & arc2 \& ICL-L1 	& 300V \\
0.2905 & 22.8 & 22.1    & 2             & 10       & ICL-L1 \& L2   	& 300V \\
0.2910 & 23.5 & 22.8    & 3             &  9       & arc1 \& 2      		& 300V \\
0.4486 & 21.3 & 20.0    & 1             & 6        & Bright2        		& 600RI \\
0.4477 & 22.3 & 21.3    & 2             & 20       & arc1           		& 300V \\
0.4148 & 23.0 & 22.3    & 2             & 7        & ICL-s1 \& s2   	& 300V \\
0.4538 & 23.1 & 22.0    & 5             & 14       & ICL-L4 \& arc2 	& 300V \\
0.5355 & 22.3 & 20.9    & 1             & 8        & arc2           		& 300V \\
0.6309 & 22.7 & 21.4    & 4             & 11       & arc2 \& 3      		& 300V \\
0.6553 & 22.3 & 21.5    & 4             & 9        & ICL-s4 \& arc2 	& 300V \\
0.6282 & 23.5 & 22.5    & 5             & 9        & arc1 \& 3      		& 300V \\
0.6267 & 23.9 & 22.9    & 4             & 10       & arc3           		& 300V \\
0.6864 & 22.6 & 21.9    & 1             & 4        & ICL-s2         		& 300V \\
0.6886 & 23.0 & 22.0    & 7             & 13       & ICL-L3 \& L4   	& 300V \\
0.6861 & 23.0 & 22.1    & 4             & 8        & ICL-s4         		& 300V \\
0.6864 & 23.5 & 22.3    & 4             & 10       & ICL-s3 \& arc2 	& 300V \\
0.6879 & 23.8 & 22.8    & 4             & 7        & ICL-s4         		& 300V \\
0.8222 & 20.7 & 20.0    & 1             & 10       & ICL-s1         		& 300V \\
0.8287 & 22.7 & 22.4    & 5             & 10       & ICL-s4 \& arc2 	& 300V \\
0.8249 & 23.2 & 22.6    & 3             & 8        & arc3           		& 300V \\
0.8823 & 23.8 & 23.4    & 4             & 3.5      & ICL-s4         		& 300V \\
0.9792 & 23.3 & 22.7    & 3             & 8        & arc3 (*)       		& 300V \\
0.9626 & 23.2 & 22.7    & 4             & 5        & ICL-L4 (*)     		& 300V \\
0.9637 & 23.4 & 23.2    & 3             & 6        & ICL-s3         		& 300V \\
0.9809 & 23.8 & 23.7    & 5             & 6        & RI             		& 600RI \\
1.0220  & 24.1 & 23.3    & 4             & 3        & ICL-s4         		& 300V \\

\hline \hline
\end{tabular}
\end{table*}

\subsection{Spectral indexes}
\label{specindex}

The 4000~\AA\ break amplitude definition used in the present paper is the `narrow' 4000~\AA\ break defined by Balogh et al. (1999)
as the flux ratio in the range 4000-4100\AA\ over 3850-3950\AA\ \citep[e.g.][]{Kauffmann:2003p5046}. The error in D(4000) is calculated 
from the spectral noise in the two passbands. The equivalent widths of [OII] and of $\rm H\delta$ were measured by using the MIDAS context 
ALICE as follows: the continuum was obtained by linear interpolation through two passbands each side of the line, a Gaussian was fitted to the 
emission or absorption line, and an integration was done over the resulting Gaussian profile above or below the continuum. The continuum and line fits, 
and the integration were done interactively on a graphic window in which the spectral region of the line was displayed. Table~\ref{integr} lists 
the wavelength ranges of the sidebands used to define the fluxes and continua.


\begin{table*}
\renewcommand{\arraystretch}{0.9}
\centering 
\caption{ Wavelength bands used in the measurement of 4000~\AA\ break amplitude, and in the determination of
the continua of the [OII] and $\rm H\delta$ indexes (equivalent widths).}
\begin{tabular}{ l c c}
Index                  &  Blue band        & Red band \\ \hline
& & \\
D(4000)                & 3850 - 3950 \AA &  4000 - 4100 \AA     \\
EQW([OII])             & 3650 - 3700 \AA &  3750 - 3780 \AA    \\
$\rm EQW(H\delta)$     & 4030 - 4070 \AA &  4130 - 4180 \AA    \\

\hline \hline
\label{integr}
\end{tabular}
\end{table*}

Uncertainties in equivalent widths were deduced from simple Monte Carlo: the values of the equivalent widths are the average of 20 
continuum determinations and best Gaussian fits to the absorption or emission lines, and the errors in equivalent widths are deduced from the Monte Carlo dispersion. 
The largest index errors are for spectra in which  $\rm H\delta$ is both in absorption and in emission. In such cases 
the emission line was removed after fitting the spectrum of an A star onto all Balmer lines to estimate the depth of $\rm H\delta$ in 
absorption, and this step was added to the Monte Carlo. The errors on indexes given in Tables of combined spectra throughout the paper are those which 
were measured on combined spectra. They do not take into account the astrophysical dispersions in the distributions of individual galaxies which were used to 
build combined spectra. Those astrophysical dispersions are given in relevant Tables concerning spectral variations.

Full observational measurement errors on indexes of individual spectra were obtained by measuring
  $\rm D(4000)$ and  $\rm EQW([OII])$ on spectra with multiple observations. Thus $17\%$ of the spectra have typical errors of $4\%$ in  D(4000)
and $10\%$ in  EQW([OII]);  $54\%$ have typical errors of  $8\%$ in D(4000) and $20\%$ in  EQW([OII]); and $14\%$ have poorer spectra with typical errors
of $16\%$ in D(4000) and $40\%$ in  EQW([OII]).

\subsection{Stellar Population Analysis}
\label{popusynthesis}

In order to study the stellar population quantitatively, we applied a modified version of the spectral population synthesis code, {\it
starlight}\footnote{http://www.starlight.ufsc.br/} \citep{CidFernandes:2004p4742, Gu:2006p5026} to fit the observed and combined spectra. 
The code does a search for the best-fitting linear combination of 45 simple stellar populations
(SSPs), 15 ages, and 3 metallicities ($0.2\,Z_\odot$,
$1\,Z_\odot$, $2.5\,Z_\odot$) provided by \cite{Bruzual:2003p4498} to match
a given observed spectrum $O_\lambda$. The model spectrum
$M_\lambda$ is:
\begin{equation}
M_\lambda(x,M_{\lambda_0},A_V,v_\star,\sigma_\star) =
M_{\lambda_0}
   \left[
   \sum_{j=1}^{N_\star} x_j b_{j,\lambda} r_\lambda
   \right]
   \otimes G(v_\star,\sigma_\star)
\end{equation}
where $b_{j,\lambda} =L_\lambda^{SSP}(t_j,Z_j) / L_{\lambda_0}^{SSP}(t_j,Z_j)$ is the spectrum of the $j^{\rm th}$
SSP normalized at $\lambda_0$, $r_\lambda = 10^{-0.4(A_\lambda - A_{\lambda_0})}$ is the reddening term, $x$ is the
population vector, $M_{\lambda_0}$ is the synthetic flux at the normalization wavelength, and $G(v_\star,\sigma_\star)$ is the
line-of-sight stellar velocity distribution modeled as a Gaussian centered at velocity $v_\star$ and broadened by $\sigma_\star$.
The match between model and observed spectra is calculated as $
\chi^2(x,M_{\lambda_0},A_V,v_\star,\sigma_\star) =
   \sum_{\lambda=1}^{N_\lambda}
   \left[
   \left(O_\lambda - M_\lambda \right) w_\lambda
   \right]^2
$, where the weight spectrum $w_\lambda$ is defined as the inverse
of the noise in $O_\lambda$. The code yields a table with input and
ouput parameters for each component. Input parameters include individual
stellar masses, ages, metallicities, L/M, ... and ouput parameters include
luminosity fractions, mass fractions, fit parameters of individual
components ..., and global parameters such as velocity dispersion and 
extinction. For more details we refer to the paper by
\cite{CidFernandes:2005p4910}. In the present work we use
the standard luminosity fraction in the rest frame
of normalized spectra at $\rm 4050~\AA$, which we compare in
different redshift bins.


 \begin{figure}
\centering
  \vspace*{0.0cm}
              \includegraphics[width=11cm]{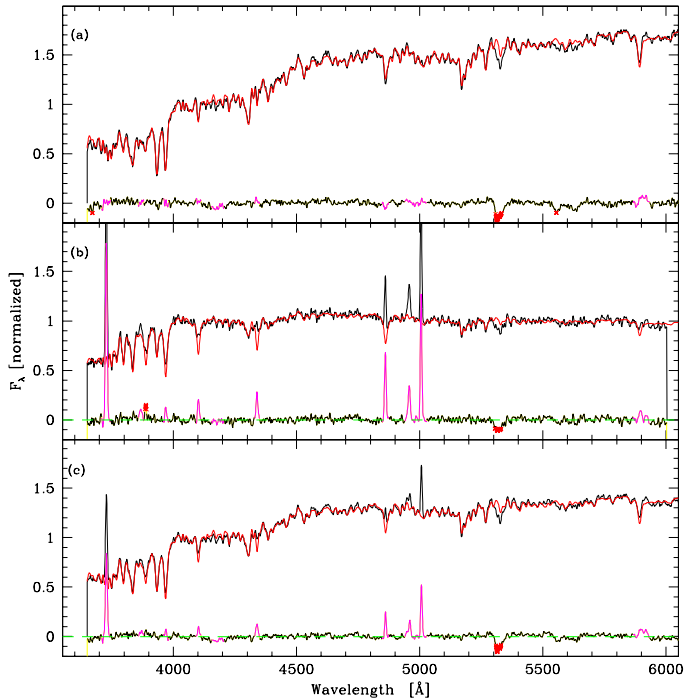}
                   \vspace*{0.0cm}
                  \caption{Spectral fitting results with SSP models for the redshift  $<z>=0.29$ bin. (a): Observed (thin black 
                  line), model (red line) and residuals for the absorption spectrum. Points indicate bad pixels 
                  and emission-line windows that were masked out during fitting. (b): Emission-line spectrum;
                  (c): Total spectrum.}
    \label{pop}
\end{figure}  

Figure~\ref{pop} shows an example of the fit for the averaged spectrum at $<z> = 0.29$. Panels (a),  (b), and  (c) correspond to  absorption-line, emission-line, and all spectra respectively.
After fitting the spectra, we rebin the 45 SSPs into 5 components according to their age: I ($10^6 \le t < 10^8$ yr), II ($10^8 \le t < 5 \times 10^8$ yr), III ($5 \times10^8 \le t < 10^9$ yr), IV ($10^9 \le t < 2.5 \times 10^{9}$ yr), and V ($t \ge 2.5 \times 10^{9}$ yr). Components with the same age and
different metallicities are combined together.


\section{The Catalog of Galaxies and Large Scale Structures in the Line of Sight in the Pencil Beam}
\label{structures}

Table~\ref{catalog} presents positions, redshifts, Petrossian R-magnitudes 
($m_R$), and line identifications for the full sample of 654 galaxies observed 
in our program. The radial velocities and the corresponding measurement errors 
are also given. The full Catalogue from which Table~\ref{catalog} is extracted
will be sent as a public database to CDS. The rough data are presently in
the public domain at ESO.


\begin{table*}
\renewcommand{\arraystretch}{0.8}
\centering
\vskip 3cm
\caption{Properties of galaxies in the field of RX J0054.0-2823 }
\begin{tabular}{|c| c| c| c| c| c| c| c| c| l|}
\hline\hline
obj & RA ($\alpha$) & DEC ($\delta)$  &  z 	&   $m_R$ 	&  V      	& V$_{err}$ & $R_T$  &  Nobs   &  lines		\\  \hline
    & J2000         & J2000           &     	&      	        & km/s 	        & km/s      &        &         &             		\\
\hline
23 & 13.598707 & -28.434965 		&  0.79304 & 22.78	& 237912 & 161 & 4.7   & 1 & K--H 					\\ 
26 & 13.590379 & -28.416917 		&  0.77636 & 22.26      & 232908 & 77   & 8.1   & 1 & [OII]--H10--H9--H 			\\  
27 & 13.584442 & -28.394515 		&  0.41463 & 22.29 	& 124389 & 22   & 17.3 & 1 & [OII]--H9--H--H$\beta$--[OIII]\\  
28 & 13.586628 & -28.438063 		&  0.44877 & 22.02	& 134631 & 73   & 6.7   & 1 & K--H--G				\\  
30 & 13.580301 & -28.435437 		&  0.29032 & 21.08	& 87096   & 68   & 11.3  & 1 & H9--K--H--H$\delta$--H$\alpha$ \\  
31 & 13.572009 & -28.380385 		&  0.63267 & 21.32	& 189801 & 49   & 11.5 & 1 & [OII]--K--H				\\  
32 & 13.579012 & -28.439414 		&  0.45335 & ---	& 136005 & 22   & 19.5 & 1 & [OII]--H$\gamma$--H$\beta$--[OIII] \\  
33 & 13.574997 & -28.439377 		&  0.44741 & 19.21	& 134223 & 80   & 11.4 & 1 & K--H--G--H$\beta$		\\  
34 & 13.573913 & -28.442855 		&  0.63013 & 20.63	& 189039 & 87   & 7.9   & 1 & K--H--H$\delta$--G		\\  
35 & 13.571781 & -28.435423 		&  0.44862 & 20.26	& 134586 & 73   & 9.8   & 1 & H9--K--H--G--H$\beta$\\
\hline\hline
\end{tabular}
\label{catalog}
\end{table*}

\begin{figure}
   \vspace*{-0.0cm}
   \includegraphics[width=11cm]{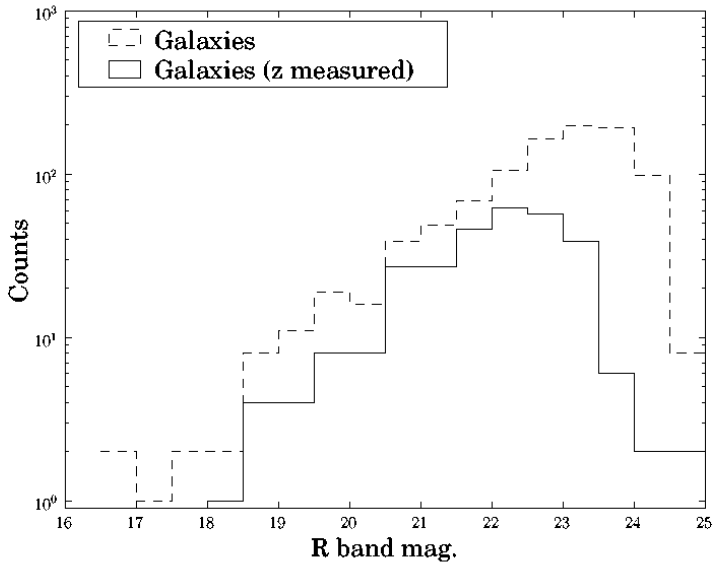}
       \vspace*{-0.0cm}
    \caption{R-magnitude histogram of galaxies with measured redshift in the central beam. }
    \label{central_R}
\end{figure}

Figure~\ref{central_R} shows the R-magnitude histogram of the galaxies with measured redshifts
superimposed on the magnitude histogram of all galaxies in our pencil-beam indicating that our observations
sample uniformly at a rate of 50-60\% the population of galaxies down to $\rm R = 22.5$. The sampling
seems fairly representative in the magnitude bin $\rm R = 22.5-23.0$, and sparse at $\rm  R > 23$. 
The apparent increase in incompleteness toward brighter magnitudes is due to a selection bias
in the observations, which were designed to avoid bright galaxies at redshifts $z\leq0.25$.


\begin{figure}
\vspace*{-0.0cm}
\includegraphics[width=15cm]{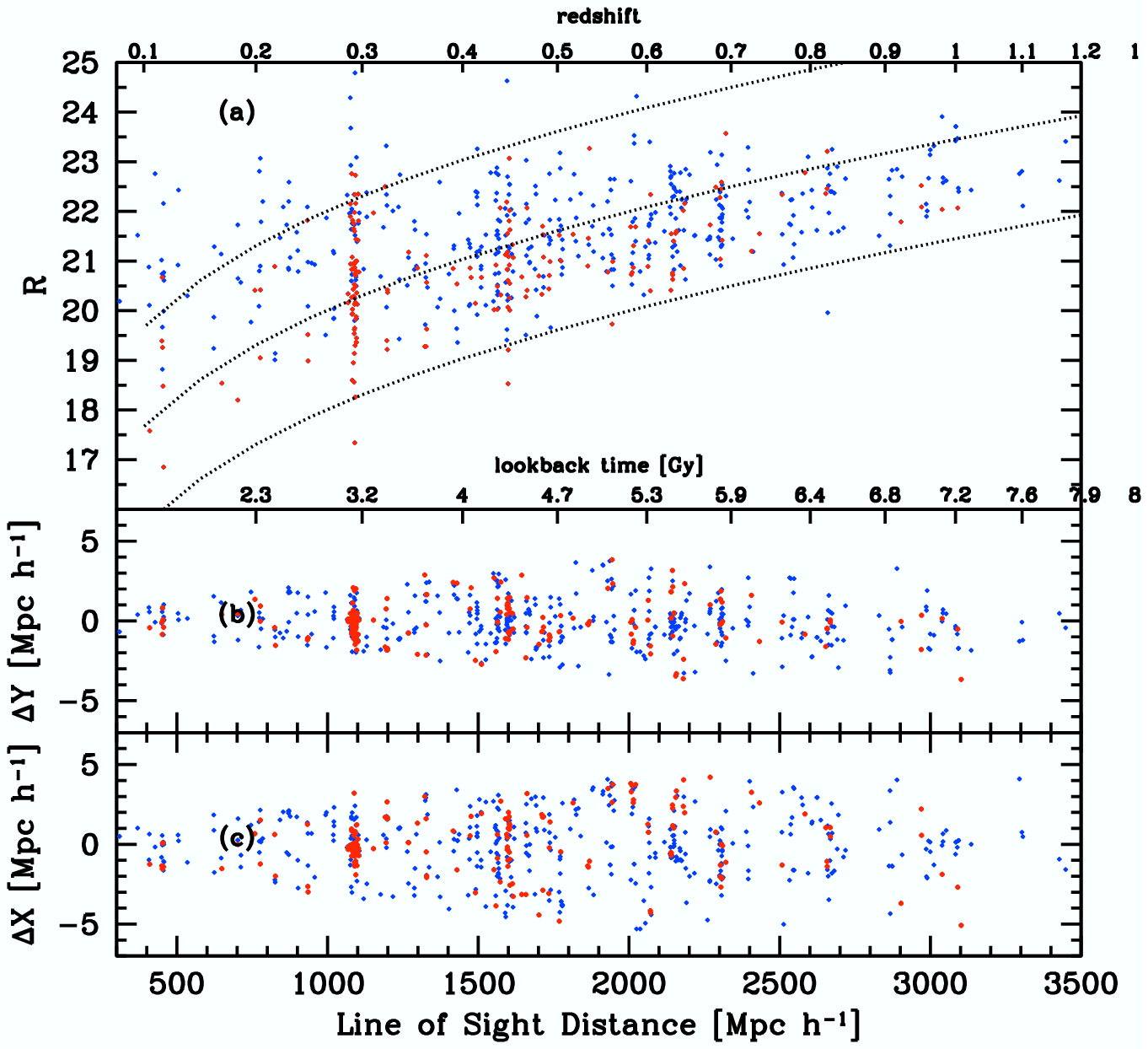}
     \vspace*{-0.0cm}
    \caption{ (a) Magnitude redshift relation for the full sample. The three lines overploted over
the measured points correspond to absolute R magnitudes of -22.5, -20.5, and -18.5.
The distances have been calculated using a cosmology with $\Omega_{0,\Lambda}=0.70$, 
$\Omega_{0,m}=0.30$, $w=-1$, and $\rm H_0 = 75 km~s^{-1}~Mpc^{-1}$ (h = $\rm H_0/75 km~s^{-1}~Mpc^{-1}$). 
Red dots are galaxies with no emission lines and blue dots are galaxies with emission lines.
(b) Cone diagrams in Dec for all the galaxies measured in the field of RX J0054.0-2823. 
The scales is in Mpc calculated using the angular distance for the standard cosmology.
The detection threshold for emission-lines is $\rm EQW([OII])\sim 2-3$\AA. 
(c) Same as (b) but for RA.
}
\label{PieDia}
\end{figure}

Figure~\ref{PieDia} presents the magnitude redshift relation and the cone diagrams for the
full sample. The points are color coded according to the presence or absence of emission lines. 
 

A cursory inspection of Figure~\ref{PieDia} reveals the presence of several conspicuous structures -
walls of objects spanning almost the entire field of view - over the full range of redshifts covered
by our observations.  Ignoring objects with $z<0.28$, we see structures centered at $z=0.29$ (our
prime target); two distinct structures at $z\sim0.4$, which we will denote $z=0.415$ and $z=0.447$;
a rather complex structure at $z\sim0.6$, with two main over-densities at 
$z=0.58-0.63$, and $z=0.68$; a single rather sparsely populated layer at $z=0.82$. In what follows,
we will refer to these groups (including the main cluster at $z=0.29$) as our pencil beam structures. 
Making bins centered on the peaks of the redshift distribution maximizes the number of objects in each bin and minimizes its redshift 
dispersion. So using the apparent structures rather than a blind slicing appears well adapted to our sample. 
If the structures are real, the objects of a given structure may have a common history and this may also help to reduce the cosmic 
scatter.

The numbers of spectra observed in each structure are given in Table~\ref{zdist}.
The redshift bins given in column (2) are chosen {\it a posteriori} to fit the structures.
The numbers of redshifts measured in each bin are given in column (3), with the respective numbers of absorption 
and emission systems in columns (4) \& (5). We have determined the median distance to the nearest object
$\triangle \delta$ in each of the apparent structures in column (6), and the median velocity dispersion,
$\rm \sigma (V) \equiv c \sigma(z) /(1+z)$ in column (7). We give a rough morphology of the structures in column (1).
The cluster at $z = 0.293$ appears to have small projected separation and velocity dispersion.
Layers or filaments have a comoving velocity  dispersion (dynamical and cosmological) less than $\rm \sim 1500~km~s^{-1}$; clouds have  
$\rm \sigma(V) > 2000~km~s^{-1}$. The structures marked ``filaments'' are the arc layers seen in Figure~\ref{PieDia}.
Projected on the sky they seem to be  filamentary, but the median distances  $\triangle \delta$ to the
nearest object are approximately 2/3 those expected for uniform distributions, so they are not 
clearly different from 2D layers. The names of the bins that are used to combine spectra are
given in column (8). Large scale arc structures, as seen in cone diagrams, 
are expected to be formed by infall of galaxies on gravitational potentials: galaxies
which are on the far side have a negative infall velocity, while those on 
the nearby side have a positive infall component, which when superimposed on the Hubble flow
reduces the velocity dispersion. This is presumably what we
observe in the two filaments or layers with low velocity dispersion at z = 0.4.

We combined the spectra in each structure using the median.
This results in a slightly lower total S/N (by $\sqrt{2}$), but allows to eliminate spurious features.


\begin{table*}[]
\renewcommand{\arraystretch}{0.9}
\centering 
\caption{Apparent structures in the field of RX J0054.0-2823 }
\begin{tabular}{c c c c c c c c }

Apparent Structure    &          z range   &     N    &   N(abs) &  N(em)  & $\triangle \delta~ (kpc)$ & $c\sigma(z)/(1+z)$        &       Composite name \\ 
   (1)      &               (2)   &     (3)  &   (4)  &    (5)  &       (6)                     & (7)  & (8)\\  \hline
& & & & & & \\
                      & 0.275 - 0.285  &  0   &    0  &    0  &               &              &      \\
Cluster               & 0.285 - 0.298  & 91   &   60  &   31  &    165        &  527         & SPEC029       \\
                      & 0.298 - 0.320   & 5    &    1  &    4  &               &              &     \\
                      & 0.320 - 0.330   & 12   &    7  &    5  &               &              &     \\
                      & 0.330 - 0.390   & 28   &    7  &   21  &               &              &      \\  
                      & 0.390 - 0.430   & 35   &    6  &   29  &    490       &   3250       & SPEC0415      \\
Filament (layer)      & 0.432 - 0.440  & 29   &    5  &   24  &     500          &   350    & SPEC0415     \\
                      & 0.440 - 0.444  &  1   &    1  &    0  &               &              &            \\
Filament (layer)      & 0.444 - 0.456  & 46   &   21  &   25  &     470      &   450        & SPEC0447      \\
                      & 0.456 - 0.465  &  1   &    1  &    0  &               &              &          \\
Cloud                 & 0.465 - 0.550   & 53   &   15  &   38  &   480        &  3370       &       \\
Cloud                 & 0.550 - 0.620   & 56   &   15  &   41  &   520        &  3550       & SPEC063         \\
Filament (layer)      & 0.620 - 0.657  & 48   &   12  &   36  &    580        &  1450      & SPEC063        \\
                      & 0.657 - 0.673  &  1   &    0  &    1  &               &              &         \\
Filament (layer)      & 0.673 - 0.696  & 43   &   12  &   31  &   650         &   1010   & SPEC068        \\
                      & 0.710 - 0.790   & 22   &    4  &   18  &               &              &            \\
Filament \& cloud     & 0.790 - 0.850   & 33   &    6  &   27  &  820         &   2050   & SPEC082       \\
                      & 0.850 - 0.880   &  0   &    0  &    0  &               &              &        \\
Cloud                 & 0.880 - 0.930   & 11   &    1  &   10  &               &              &   \\
                      & 0.930 - 0.946  &  0   &    0  &    0  &               &              &      \\
Cloud                 & 0.946 - 1.046  & 25   &    4  &   21  &    1080    &    4110   & SPEC099        \\

 \hline \hline
\label{zdist}
\end{tabular}
\end{table*}

\subsection{Magnitudes}

The R-band average magnitudes of galaxies in each redshift bin are given in Table~\ref{magnitudes} separately for absorption, ``red'' and  ``blue'' emission-line galaxies,
together with the adopted distance moduli. The partition ``red'' versus ``blue'' is defined by the median spectral slope in each redshift bin. In a study
on emission line galaxies \citep{Giraud:2010} we divided the sample of 
emission-line galaxies in two halves: those with continuum slopes  bluer than 
the average  and those with continuum slopes redder than the average in each redshift bin. This was done interactively by displaying reduced 1D spectra
and using MIDAS. While a  median partition is not necessarily a physical 
partition, we showed that, in the present case, it divides "young" galaxies, 
for which the evolution is dominated by on-going star formation from "old" 
galaxies where the evolution is dominated by changes in the older stellar 
populations.


\begin{table*}
\renewcommand{\arraystretch}{0.9}
\centering 
\caption{Average R-band  magnitudes of absorption systems (abs), and  red and blue emission-line galaxies. The adopted distance moduli $(m-M)_0$ and the 4150-4250\AA\  fluxes $f$ normalized to the blue galaxies at $z = 0.9$  are also tabulated.}
\begin{tabular}{ c c c c c c c c}
$<z>$     & R(abs)  & R(red) & R(blue) & $(m-M)_0$  &     $f$(abs) & $f$(red)  & $f$(blue)  \\ \hline
& & & & & & & \\
0.29         & 19.80    & 20.12  & 20.97     & 40.18   &   0.74     & 0.72    & 0.48   \\ 
$0.43$    & 20.24    & 20.59  & 20.95     & 40.86   &   1.08     & 0.92    & 0.75    \\
$0.65$    & 21.50    & 21.60  & 21.94     & 41.51   &   1.42     & 1.28    & 0.86   \\
$0.9$      & 22.45    & 22.13  & 22.35     & 41.98    &   2.08    & 1.70    & 1  \\
\hline \hline
\label{magnitudes}
\end{tabular}
\end{table*}

We used the R-band photometry to calibrate individual spectra by convolving each spectrum with a box filter 1290~\AA\ wide, centered 
at $\lambda=6460$~\AA. Once the spectra were calibrated in the observer R-band, we measured the average fluxes in the wavelength range  
4150-4250\AA\ of the galaxies, which we normalized to the flux of blue emission galaxies at $<z> = 0.9$ to compute the luminosity index $f$. Thus $f$ (that is equal to 1 for blue 
galaxies at $<z> = 0.9$) is an indicator of AB(4200) that allows us to compare the luminosities of 
red and blue galaxies at a given redshift and to investigate luminosity variations with $z$. Thus Table~\ref{magnitudes} clearly shows that in each redshift bin, absorption-line and red emission-line galaxies are more luminous than
blue galaxies.


\section{Composite spectra}
\label{evolution}

Each galaxy spectrum was wavelength calibrated, corrected for instrument response,
re-binned to zero redshift, and normalized to have the same flux in the wavelength range 
$\rm \triangle \lambda=4050-4250\AA$. Normalizing spectra gives the same
weight to all galaxies. As a consequence stellar fractions must be understood 
as average stellar fractions per galaxy. 

\begin{figure}
  \vspace{1mm} 
 \centering
      \includegraphics[width=11cm]{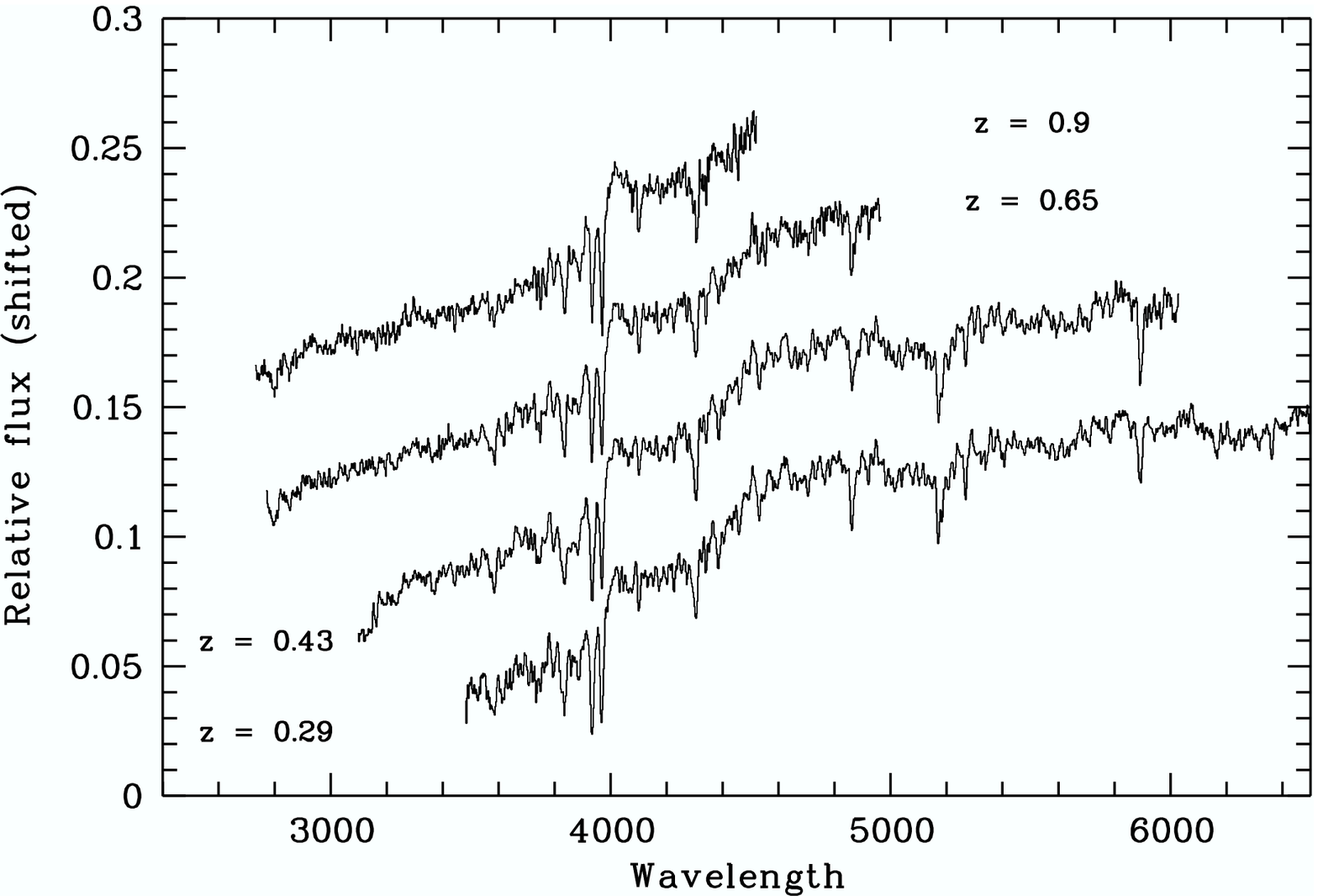}
\\
      \includegraphics[width=11cm]{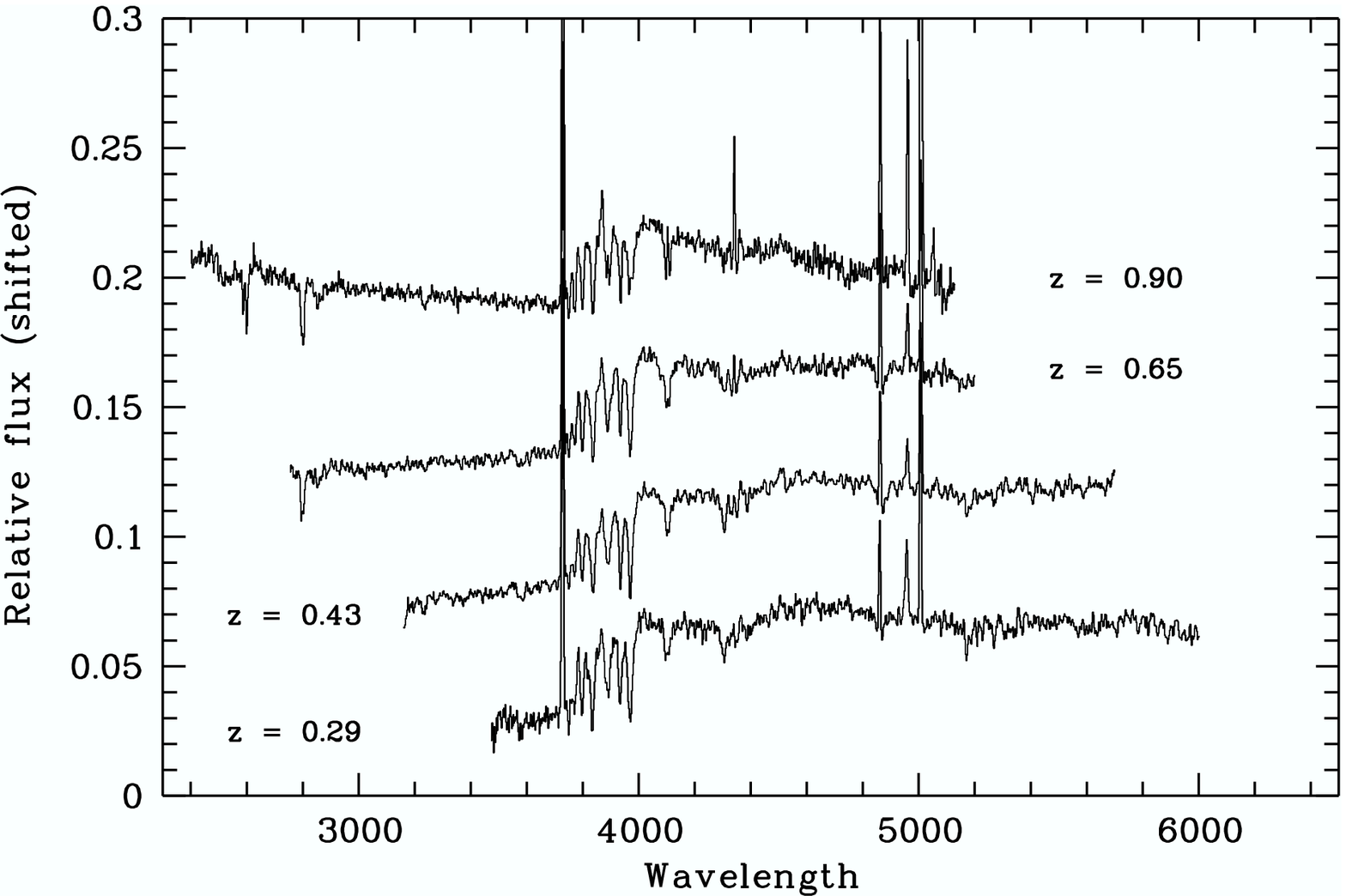}
  \vspace{-0.0cm}
    \caption{Composite spectra of absorption systems (top); and emission line galaxies (bottom) normalized
		in the wavelength range $\rm \triangle \lambda = 4050 - 4250$~\AA. All individual galaxies are brighter than 
$\rm M_R = -18.8$}
    \label{All}
\end{figure}

We have truncated the sample at $z = 1.05$ and assembled the spectra in bins
centered on (pseudo) structures at  0.29, 0.41, 0.45, 0.63, 0.68, 0.82, and 0.99 to build high S/N composite spectra for each bin. 
In order to compensate (or at least alleviate) for Malmquist bias we rejected objects fainter than $\rm M_R = -18.8$ mostly at $z \leq 0.45$ (Figure~\ref{PieDia}a). A sample completely free of Malmquist bias would require a cutoff at  $\rm M_R \sim -20.5$. For clarity of the figures, we often combined  
the mean spectra at  $z=0.41$ \& $0.45$ into a single bin at  $<z>=0.43$, the spectra 
at $z=0.63$ \& $0.68$ into a bin at  $<z>=0.65$, and in some cases the spectra at $z=0.82$ \& $0.99$ into a bin at  $<z>=0.9$.
The spectra of galaxies in these four bins are presented in  Figure~\ref{All} where we show the
spectra of absorption systems (top) and emission line galaxies (bottom) separately. The corresponding \rm $4000 \AA$ break amplitudes
are given in Table~\ref{D4000} \rm 

%


\begin{table*}
\renewcommand{\arraystretch}{0.9}
\centering 
\caption{  4000\AA\  break amplitudes for absorption (abs) and emission (em) galaxies,
and equivalent width of H$\delta$ for absorption galaxies with measurement errors.
The S/N ratios of the combined spectra were measured in the interval  4050\AA--4250\AA. The magnitude cutoff is
 $\rm M_R = -18.8$ for all redshift bins.}
\begin{tabular}{  c c c c c c}
      & \multicolumn{3}{c}{Absorption systems} & \multicolumn{2}{c}{Emission systems} \\
$<z>$ & D(4000)           & $\rm EQW(H\delta)$       & S/N  & D(4000)          & S/N   \\ \hline
& & & & \\
 0.29    & $1.67 \pm 0.065$  &  $-1.5 \pm 0.2$          & 23   &  $1.22 \pm 0.02$ & 32       \\
$ 0.43$  & $1.70 \pm 0.06$   &  $-1.5 \pm 0.2$          & 22   &  $1.22 \pm 0.01$ & 52      \\
$0.65$   & $1.60 \pm 0.055$  &  $-1.8 \pm 0.2$          & 24   &  $1.14 \pm 0.01$ & 35      \\
0.82     & $1.57 \pm 0.06$   &  $-2.4 \pm 0.5$          & 18   &  $1.07 \pm 0.02$ & 28      \\     
0.99     & $1.43 \pm 0.05$   &  $-2.9 \pm 0.3$          & 23   &  $1.08 \pm 0.02$ & 25      \\
\hline \hline
\label{D4000}
\end{tabular}
\end{table*}

The most conspicuous spectral change with redshift is a decrease in flux redward of the G-band from $<z>=0.29$ 
and $<z>=0.43$ to higher $z$ coupled to an increase to the blue of [OII] from  $<z>=0.65$ to  $<z>=0.82$ and higher $z$
in emission-line galaxies. This  systematic change of the continuum implies
that the galaxy population varies  as a function of redshift: more star forming galaxies at higher $z$
and more galaxies with old stars at  lower $z$. This spectral change, which is known, will not be studied further in this paper except to quantify (in~\ref{linerimpact})
the impact of LINER-like galaxies at $z=0.4--0.9$. In the following section we concentrate
on absorption systems and low-ionization galaxies.


\section{Absorption line systems}
\label{absorption}

The spectral resolution of the 300V grism allows us to detect  [OII] emission down to  $\rm EQW([OII]) \sim 2-3~\AA$. We will call
absorption-line galaxies those for which any mechanism of ionization is low 
enough to preclude [OII] detection at our detection level. Thus, our pure
absorption-line sample comprises mostly E, E+A, and S0 galaxies  with no 
on-going star formation, nuclear activity, or other mechanism of ionization.

\subsection{Absorption systems as function of redshift}

The normalized and combined spectra of absorption line systems presented in Figure~\ref{All} (top) do not show any obvious change in their continuum and 4000\AA\ break amplitude up to 
$z \approx 0.6$ (Table~\ref{D4000}). There is a moderate decrease in the  4000\AA\ break at $z \geq 0.65$ ranging from  $5\%$ at $z \sim 0.65$ to $7\%$ at $z \sim 0.82$ 
and up to $15 \%$ at $z \sim 1$, while the
$\rm H \delta$ absorption line becomes stronger at $z \geq 0.65$ (Table~\ref{D4000}), suggesting 
the presence of increasing numbers of A stars at higher redshifts.  The indexes suggest that these galaxies 
had the bulk of their star formation at $z \geq 1$, while some of the systems at $z > 0.8$ still had clearly 
detectable star formation about 1 Gyr ago.

We have compared our spectral indexes at  $z \sim 0.82$ with those
measured by \cite{Tran:2007p5181} in the rich cluster MS 1054-03 at z = 0.83 using the same index
definitions from \cite{Kauffmann:2003p5046}. The average break amplitude and $\rm H\delta$ index of  
absorption systems in MS 1054-03 are respectively  $\rm D(4000) (abs) = 1.67 \pm 0.00$ and 
$\rm EQW(H\delta) (abs) = -1.7 \pm 0.0$ \citep[][Table 4]{Tran:2007p5181}. Our absorption systems at 
$ z \sim 0.82$ appear to have younger stellar populations  as indicated both by $\rm D(4000)$ and $\rm EQW(H\delta)$ (Table~\ref{D4000}). Therefore our  
absorption systems contain A stars, but clearly less than composite  field E+A  galaxies at $<z> = 0.6$  for which $\rm D(4000) (abs)=1.36\pm0.02$ and 
$\rm EQW(H\delta) (abs)=-4.6\pm0.2$ (quoted in Tran et al. (2007, Table 4) 
from data in Tran et al. (2004)). Consequently our average spectrum at  $z \sim 0.82$ is intermediate 
between pure E and pure E + A. In fact, our SSP models (Table~\ref{poptable})  indicate that absorption-line systems at $z\geq0.65$ contain on average more than $50 \%$ of stars younger than 2.5Gyr per galaxy, while those at $z \geq 0.8$ had significant star formation as recently as one Gyr ago (Table~\ref{poptable}). 


\begin{table*}
\renewcommand{\arraystretch}{0.9}
\centering 
\caption{Stellar population properties of normalized average absorption (abs) spectra in
each redshift bin. The magnitude cutoff is at $\rm M_R = -18.8$, except for the 10 faintest absorption systems at $ z = 0.29$ where
we used all the observed objects. The fractions indicated in all SSP Tables
are standard luminosity fractions at $\rm 4050 \AA$, as in Cid-Fernandes et 
al. (2010, and references therein)}
\begin{tabular}{ c l c c c c c}
  $<z>$   &  \ \ \ \ \ \ \ \ \ \ log(Age):  &    $<8$  & $8-8.7$ & $8.7-9$ & $9-9.4$& $>9.4$\\ \hline
& & & & & & \\

$0.29$                        & abs            	&  0.0\%        & 0.0\%    	&  30.1\%       &  0.1\%  	&  69.8\% \\
         		& abs (10 brightest)    &  0        	& 0      	&  17.4   	&  0    	&  82.4 \\
         		& abs (10 faintest)  	&  0        	& 0      	&  12.0   	&  66.4 	&  21.7 \\

$0.43$                        & abs                   &  0.0          & 0.7           &  11.7         &  6.9          &  80.7 \\

$0.65$                        & abs                   &  0.0          &  0.0          & 18.2          & 38.5          & 43.3 \\

$0.82$                        & abs                   &   0.0         &  0.0          & 86.8          &  3.3          &  9.8\\

$0.99$                       & abs                   &  0.0          & 0.0           & 42.3          & 0.0           & 57.7 \\

\hline \hline
\label{poptable}
\end{tabular}
\end{table*}

Post-starburst E+A galaxies are thought to be in a transition phase between a star-forming period and a 
passively evolving period. Being close to the phase of shutdown or {\it quenching} of star formation, they probably  
play an important role in the build-up of early-type systems \citep[e.g.][]{Wild:2009, Yan:2009}.
Studies of intermediate redshift clusters at $0.3\leq z\leq0.6$ have found either a higher fraction of 
post-starburst galaxies in clusters than in the field \citep{Dressler:1999, Tran:2003, Tran:2004p5213}, or a similar fraction \citep{Balogh:1999p4259}.
In fact, there is a strong variation in the E+A fraction between the SSDS low redshift survey at $z\sim0.07-0.09$,
and high $z$ surveys at $z \approx 0.5-1$ \citep[VVDS,][]{Wild:2009}, or $z \approx 0.7-0.9$ 
\citep[DEEP2,][]{Yan:2009}. 

\begin{figure}
   \centering
   \vspace*{-.0cm}
   \includegraphics[width=11cm]{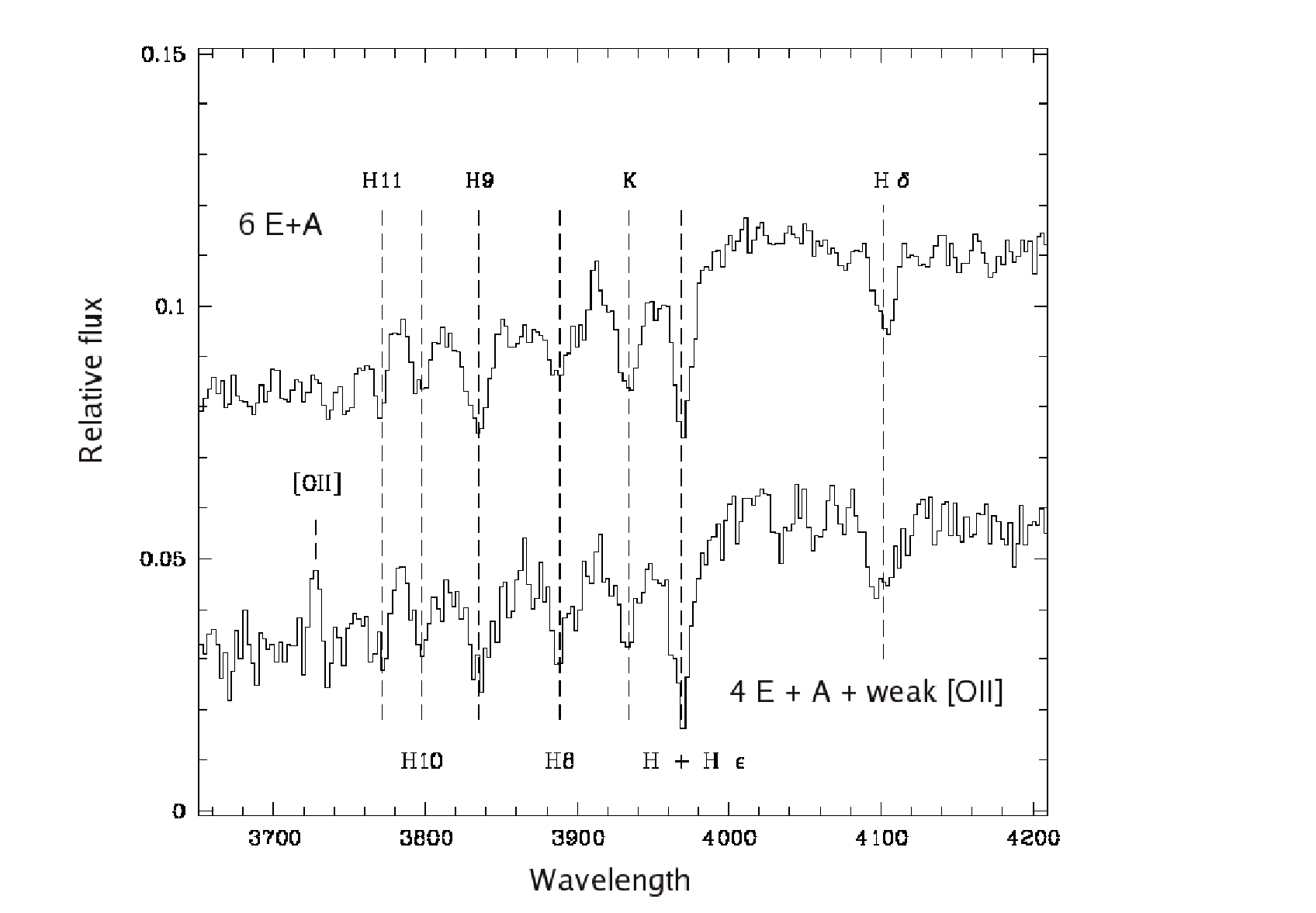}
    \vspace*{-.0cm}
    \caption{(Top) Average spectrum of 6 E+A galaxies at $0.68 \leq z \leq 1$. The Balmer series $\rm H\delta$, $\rm H+H\epsilon$,
H8, H9, H10, and H11 is very promintent and $\rm <D(4000)> = 1.40$. (Botton) Average spectrum of 4 galaxies in the intermediate redshift range 
$0.4\leq z\leq0.5$ with E/S0 morphological type, showing a poststarburst E+A spectrum with still some star formation.  $\rm <EQW([OII])> \approx 3.5 \AA$ and 
$\rm <D(4000)>=1.41$. Probably this spectrum would  have been classified as E+A at lower resolution.
}
    \label{EAOII}
\end{figure}

In order to search for  E+A galaxies in our sample we built template spectra by combining a pure
E spectrum from our sample with various fractions of an A stellar template.  We then compared our models
with absorption-line systems  in the range $\rm 1.2\leq D(4000)\leq1.5$ assuming, by definition,
that E+A galaxies contain at least $\rm 25\%$ A stars. Using this (standard) definition we searched our sample at  $0.35\leq z\leq1$ and found only 6 bona-fide E+A galaxies. 
In fact all the objects were found at $0.68 \leq z \leq 1$,  which makes our small number consistent with the VVDS and the DDEP2 surveys within a factor of 2.
The median of the normalized spectra of these 6  (as far as we can judge from our images) elliptical galaxies is presented in Figure~\ref{EAOII} (a).  We were surprised to find no E+As  at $z\sim0.4$, but we did find 4 objects with early-type morphology and very small $\rm <EQW([OII])> \approx 3.5 \AA$,  which probably would have been classified as E+As 
on lower resolution spectra. The average spectrum of these 4 objects is shown in Figure~\ref{EAOII} (b).  Their $\rm R\sim22$ magnitudes place them at the faint end of absorption line systems at the corresponding redshifts. \rm


\subsection{Absorption-line galaxies as function of luminosity at $z = 0.29$}

\begin{figure}[!]
\centering
\vspace*{0.0cm}
\includegraphics[width=11cm]{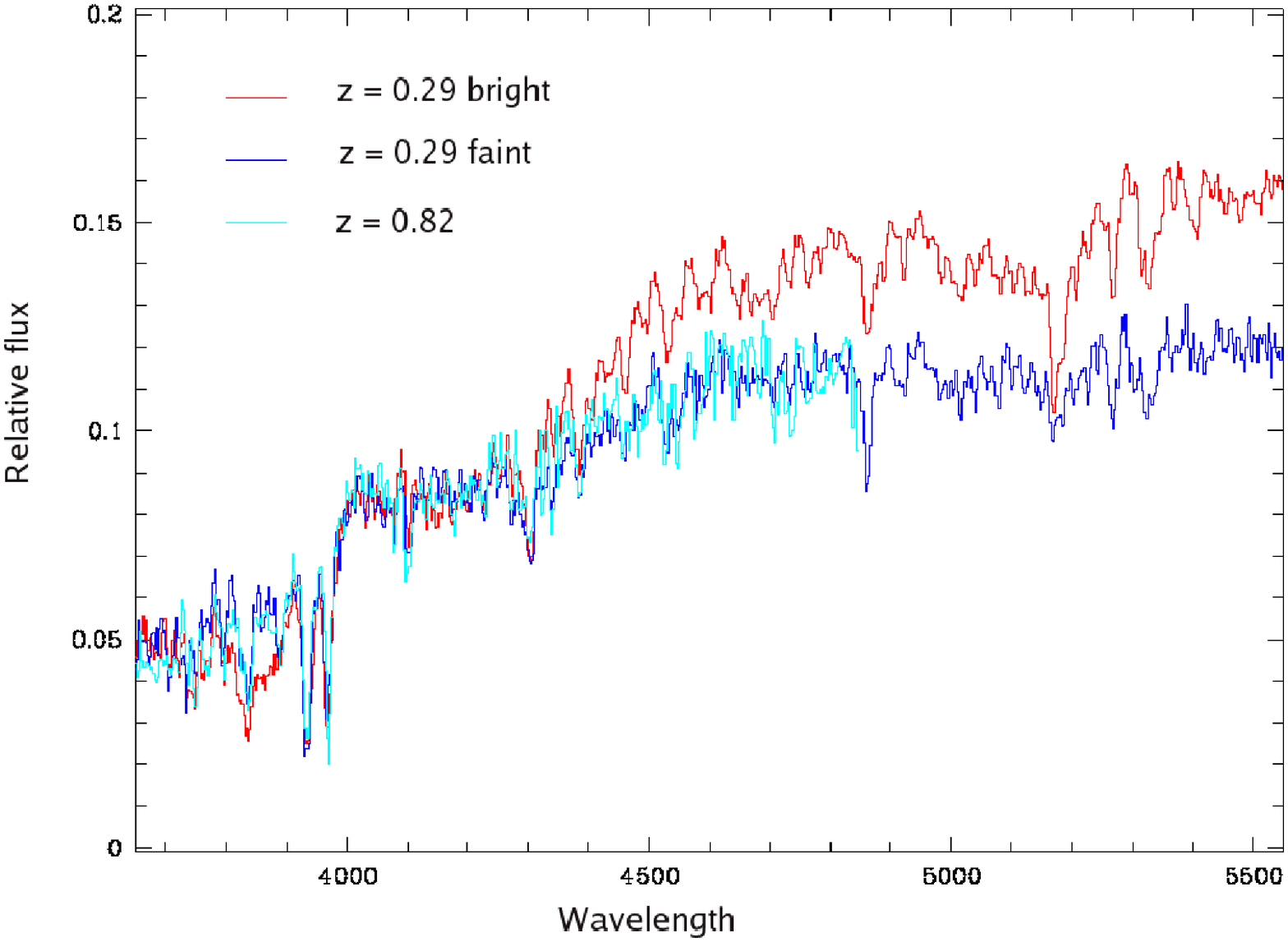}
\vspace*{-0.0cm}
\caption{Normalized spectra of the 10 brightest (in red) and the 10 faintest 
(in blue) absorption-line galaxies in the cluster at $z = 0.29$, 
and the full sample of absorption-line systems at $z = 0.82$ (in cyan).  }
    \label{SPEC029_082A}
\end{figure}

Having tested bright absorption galaxies at various redshifts (with cut-off at $\rm M_R = -18.8$), we now turn to faint 
absorption galaxies in the cluster at $<z> = 0.29$ by combining the spectra of the 10 faintest galaxies without 
emission lines. Their average R-band magnitude is $\rm R = 22$, which at a
distance modulus of 40.18 corresponds to $\rm M_R = -18.2$, and the faintest object has $\rm M_R = -17.44$.
Their mean indexes, $\rm D(4000)=1.55\pm0.01$;
$\rm H\delta=-2.27\pm0.04$, measured on the spectrum shown in Figure~\ref{SPEC029_082A}, \rm are consistent with a younger age 
than absorption-line galaxies with $\rm M_R \leq-18.8$ (Table~\ref{D4000}) in the same cluster. This is in agreement
with the well known evidence that the stellar populations in absorption systems tend
to be younger in low mass galaxies than in the more massive ones \citep[e.g.][]{Renzini:2006p5117}.
The index values are in fact very close  to those of our absorption systems at z = 0.8 (Table~\ref{D4000}) which, by selection effects, are bright  (Table~\ref{magnitudes} and Figure~\ref{PieDia}).

The SSP models indicate that on average about 80\% of the stars in the 10 faintest galaxies are younger than 2.5 Gyr (Table~\ref{poptable}), i.e. were born at $z < 1$. In comparison, 80\% of the stars contributing to the spectrum of the brightest absorption galaxies in the cluster are older than 2.5 Gyr (Table~\ref{poptable}).
To illustrate the spectral differences between bright and faint systems at $z = 0.29$, and the striking similarity between the spectra of faint galaxies at $z=0.29$ and those of bright galaxies at $z=0.8$, we have plotted in Figure~\ref{SPEC029_082A} the average spectra of the 10 brightest and the 10 faintest absorption systems at $z = 0.29$, and the average spectrum of absorption galaxies at $z = 0.82$.  The effect of {\it downsizing}, (in the present case the so-called
'archeological dowsizing') where star formation shifts from high mass galaxies at high redshifts, to low mass galaxies at low redshifts is clearly exemplified in this figure.



At a redshift of $z\sim 0.8$ (i.e. $\sim4$ Gy earlier), the red-sequence of our unrelaxed  (merging central system; elongated intra-cluster light and galaxy distribution) cluster at z=$0.29$ was already in place, but was truncated at brighter magnitudes because the faint absorption-line galaxies were still copiously forming stars. This seems consistent with the observation that some clusters at $z\simeq1$ have red sequences truncated at faint limits \citep{Kodama:2004p5056, Koyama:2007p5060}, and supports 
the picture of an environmental dependence of red-sequence truncation presented by Tanaka et al. (2005).
This is also in agreement with scenarios where the final assembling of the red-sequence can be observed
well below $z=1$ \citep{Faber:2007}.  

As discussed above, the strict definition of E+A galaxy requires a mix of an E-type spectrum with at least 25\% A stars and no traces of star formation, which in our sample implies no emission lines with equivalent widths larger than 2-3\AA. With this definition our $z=0.29$ cluster contains only one E+A galaxy while the dense layers at $z\sim0.4$, where the red-sequence is already in place (layer in Figure~\ref{PieDia}), contains none. However, both in the cluster and in the intermediate redshift layers we find plenty of galaxies with early type morphologies, A stars, and very weak emission lines. In the next section we present a closer look at these low-ionization emission line galaxies.


\subsection{Galaxy evolution and low-ionization emission-line galaxies (LINERs).}
\label{liners}

\begin{figure}
   \centering
   \vspace*{0.5cm}
   \includegraphics[width=11cm]{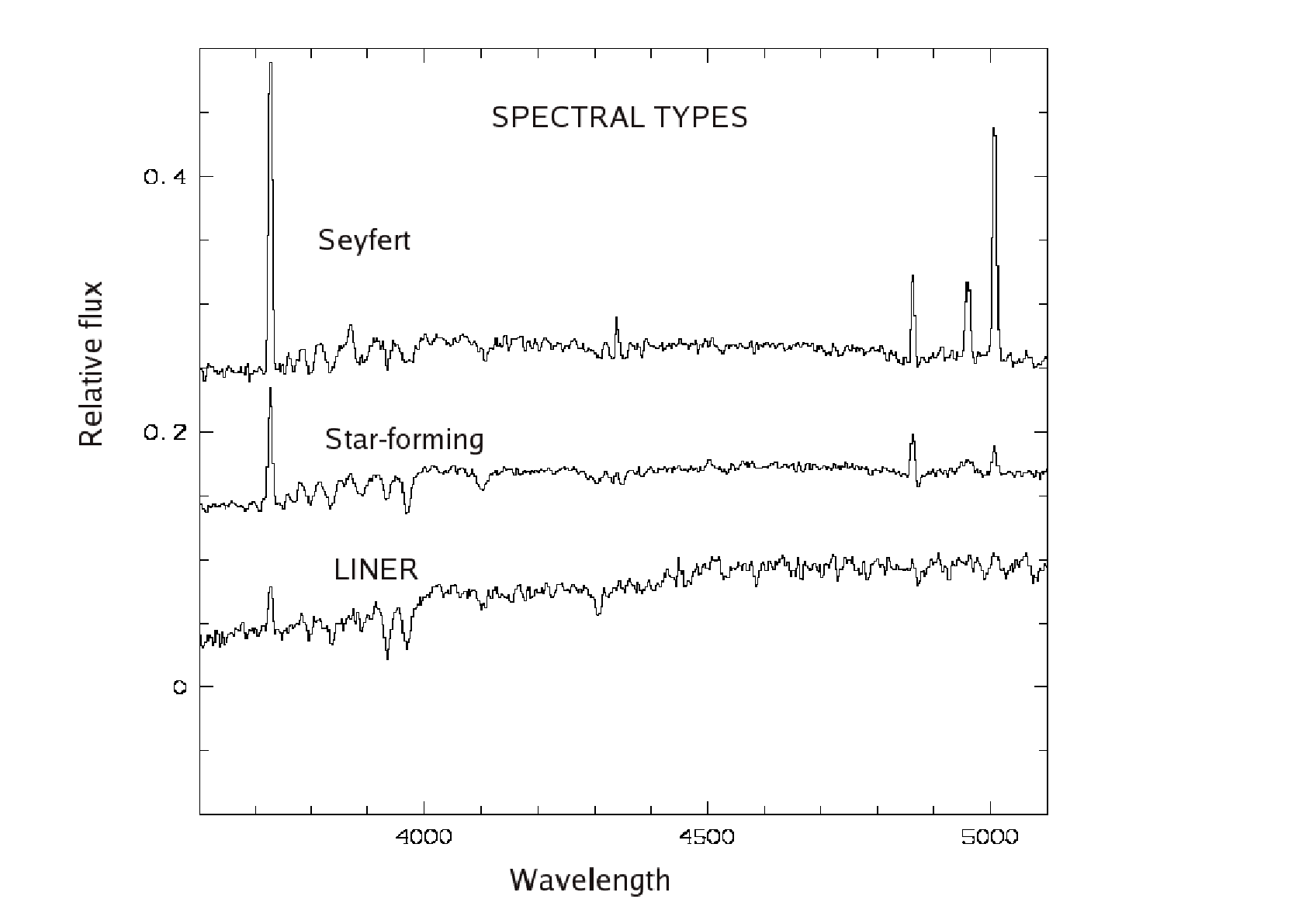}
    \vspace*{-.0cm}
    \caption{Typical average spectra of low-ionization objects, star-forming galaxies and Seyferts from the sample in the $<z> = 0.415$ layer.
}
    \label{SPEC0415lineseyfsf}
\end{figure}

In an extensive work based on the SDSS survey, Yan et al. (2006) determined the extent to which [OII] emission produced by mechanisms other than recent star formation introduces biases in galaxy evolution studies based upon [OII] only. They showed that the $\rm [OII]/H\beta$ ratio separates LINERs from star-forming galaxies, while $\rm [OIII]/[OII]$ and $\rm [OIII]/H\beta$ separate Seyferts from LINERs and star-forming galaxies. Using the classification scheme of Yan et al. (2006) we divided our spectra in 3 main classes:  LINERs, with clearly detected [OII], but no ($3 \sigma$) detection of [OIII] and $\rm H\beta$ in emission after subtracting an E+A profile;  Seyferts, with $\rm [OIII]/H\beta \geq 3$; and star-forming galaxies, which are the objects with clearly detected [OII] that are neither Seyferts nor LINERs after subtracting an E+A profile. Typical spectra of low-ionization objects, star-forming galaxies and Seyferts are shown in Figure~\ref{SPEC0415lineseyfsf}. 


\subsubsection{The impact of LINERs in our previous results on red emission-line galaxies}
\label{linerimpact}

\begin{figure}
   \centering
   \vspace*{-0.0cm}
   \includegraphics[width=11cm]{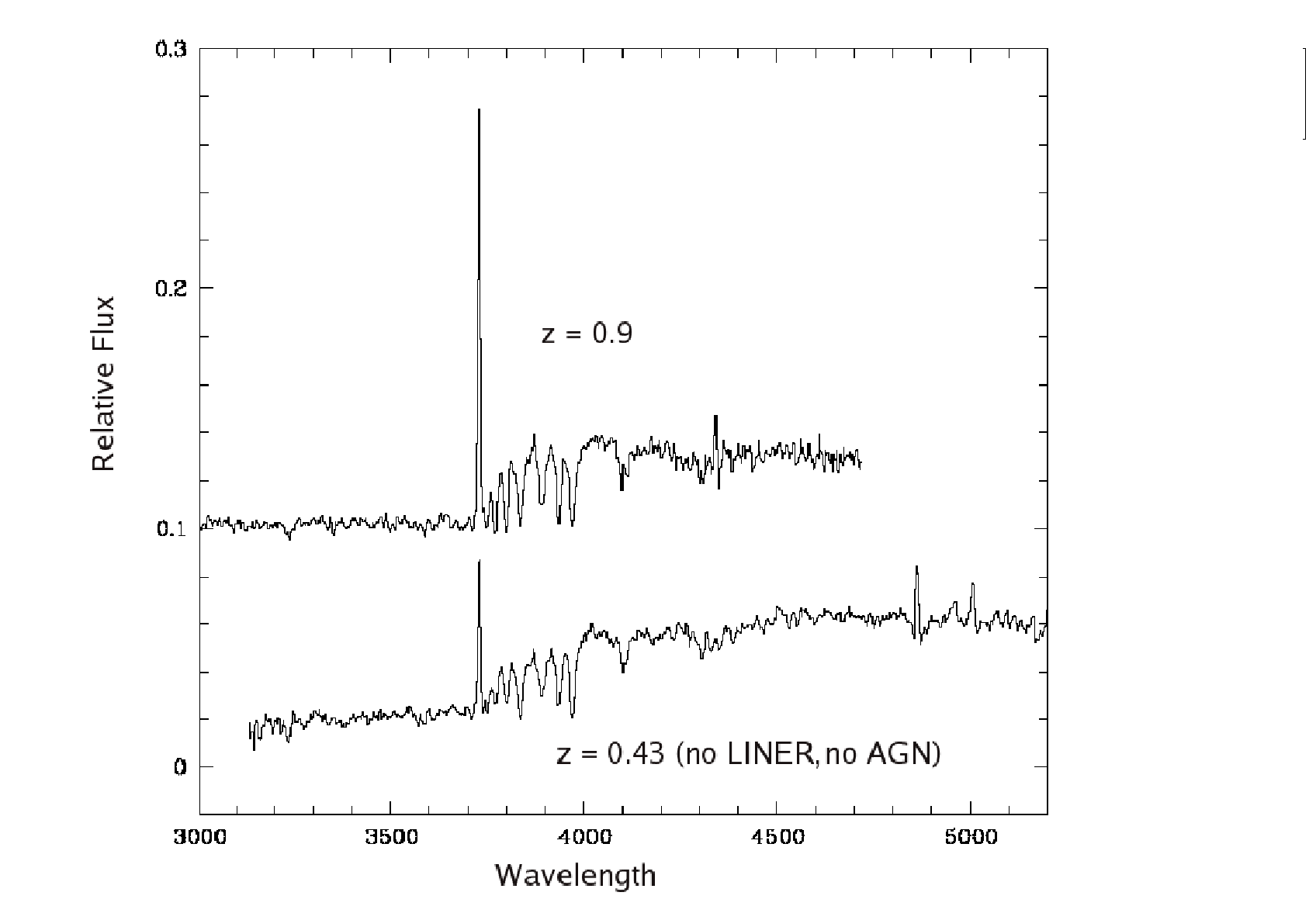}
    \vspace*{-.0cm}
    \caption{Average spectra of red emission-line galaxies after subtracting early-type LINERs and galaxies with diluted star formation
(and rare Seyferts) from the sample in the $<z> = 0.43$ bin and recalculating the median blue-to-red partition, and of the red half of 
emission-line galaxies at $<z> = 0.9$. The spectra at $<z> = 0.9$ were not classified because $\rm H\beta$ and [OIII] are missing in most cases.
}
    \label{SPEC09_043woLI}
\end{figure}

Because the spectral coverage in a rather large fraction of our objects at $<z>=0.68$ and  higher is truncated below $\rm 5000 \AA$ in the rest frame, we applied our classification scheme only to objects in the range $0.29\leq z\leq0.65$.  To extract $\rm H\beta$ in emission we built a series of E+A models, combining an observed E spectrum with different fractions of an A stellar template, ranging form $0.05\%$ to $80\%$ of the total luminosity. To determine the best-fit model we minimized the continuum slope of the difference between the spectrum and the E+A model. Thus, in the range $0.35-0.55$ our sample contains 23\% LINERs, 51\% star-forming galaxies, 8\% Seyferts, and 14\% uncertain types. The layer at $z=0.63$ has 18\% LINERs, 50\% star-forming galaxies, 7\% Seyferts, 13\% uncertain types and  11\% of truncated spectra. Altogether, the fraction of LINERs  among emission-line galaxies up to $z=0.65$ in our pencil beam is $\rm \approx 22\%$. With an average  $\rm <D(4000)>=1.39\pm0.18$. LINERs at $z\leq0.65$ have a potentially significant impact on the conclusions of Giraud et al. (2010) about the evolution of red emission-line galaxies. To quantify this impact, we have subtracted all LINERs from the sample of emission-line spectra in the $z = 0.43$ bin, determined the new blue-to-red partition (as in Giraud et al. 2010; section 5.1), and computed a new average spectrum for the red galaxies. This (also cleaned of rare red Seyferts) is shown in 
Figure~\ref{SPEC09_043woLI} where it is compared with the mean red spectrum at $z = 0.9$. We find that the differences in continuum slope and D(4000) between $<z> = 0.43$ and $<z> = 0.9$ is reduced by a factor of 2/3.
The main difference between red galaxies with LINERs and those without is the presence of young stellar population.


\subsubsection{Early-type LINERs}

The fraction of nearby early-type galaxies hosting bona-fide (i.e. nuclear) LINERs in the Palomar survey \citep{Filippenko:1985, Ho:1997a} was found to be $\sim 30 \%$ \citep{Ho:1997b}, but
LINER-like emission line ratios are also observed in extended regions \citep[][ and references therein] {Phillips:1986, Goudfrooij:1994, Zeilinger:1996, Sarzi:2006}. A similar fraction of LINER-like ratios is found in the SDSS at $0.05 \leq z \leq 0.1$ in color-selected red galaxies \citep{Yan:2006}. 

\begin{figure}
   \centering
   \vspace*{-0.0cm}
   \includegraphics[width=11cm]{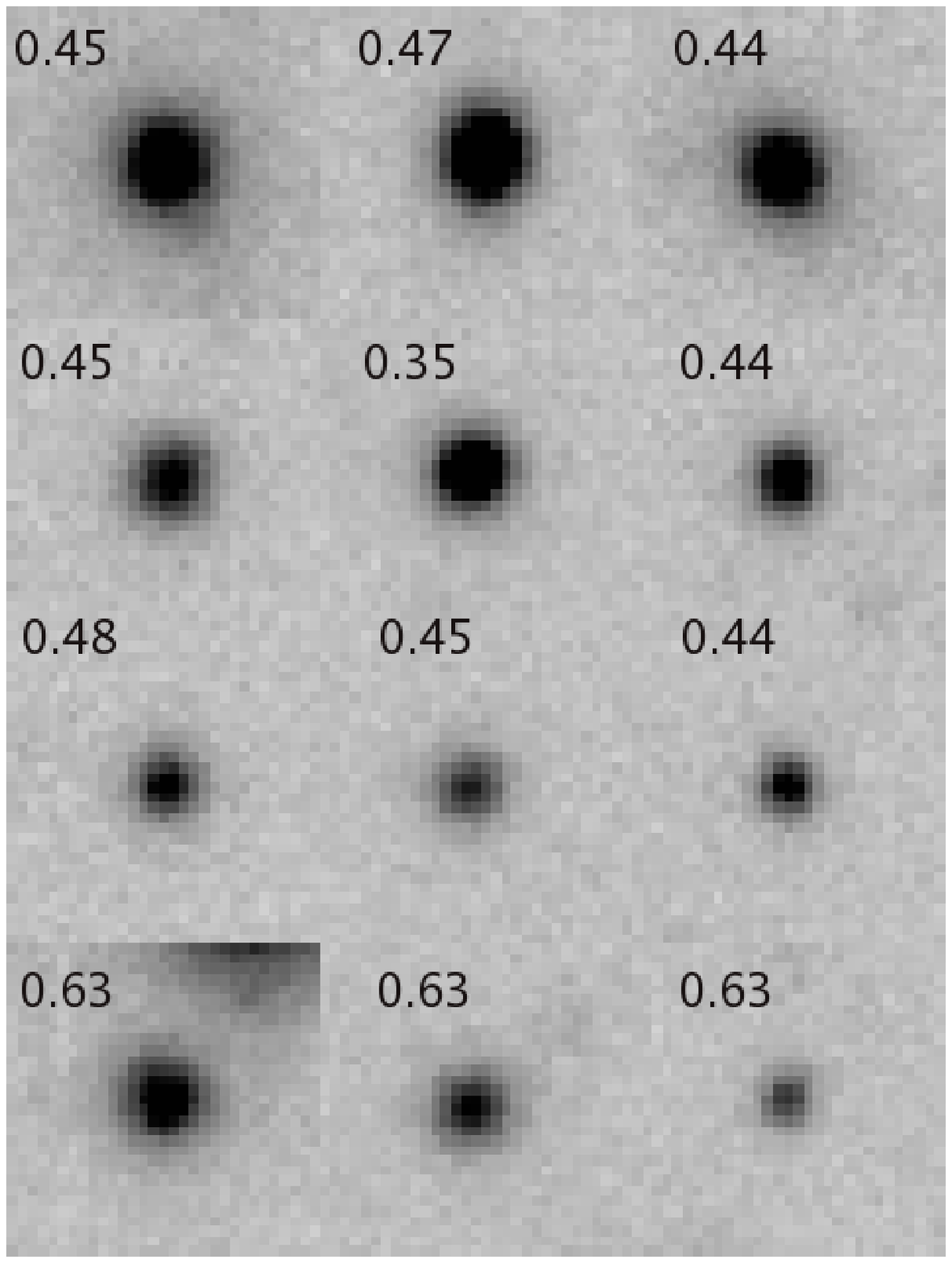}
    \vspace*{-.0cm}
    \caption{Examples of early-type galaxies having low-ionization spectra, and indicated redshifts.
}
    \label{Earlyima}
\end{figure}

Because it is very difficult to disentangle early-type LINERs from spirals with extended and diluted star formation by using only [OII] and $\rm H\beta$, we make use of morphology to distinguish compact objects with low ellipticity and profiles consistent with early type galaxies, from other morphologies: apparent disks, high ellipticity, and irregular or distorted morphologies. Images of early-type galaxies with low ionization spectra are shown in Figure~\ref{Earlyima}. 


\begin{figure}
   \centering
   \vspace*{0.5cm}
   \includegraphics[width=11cm]{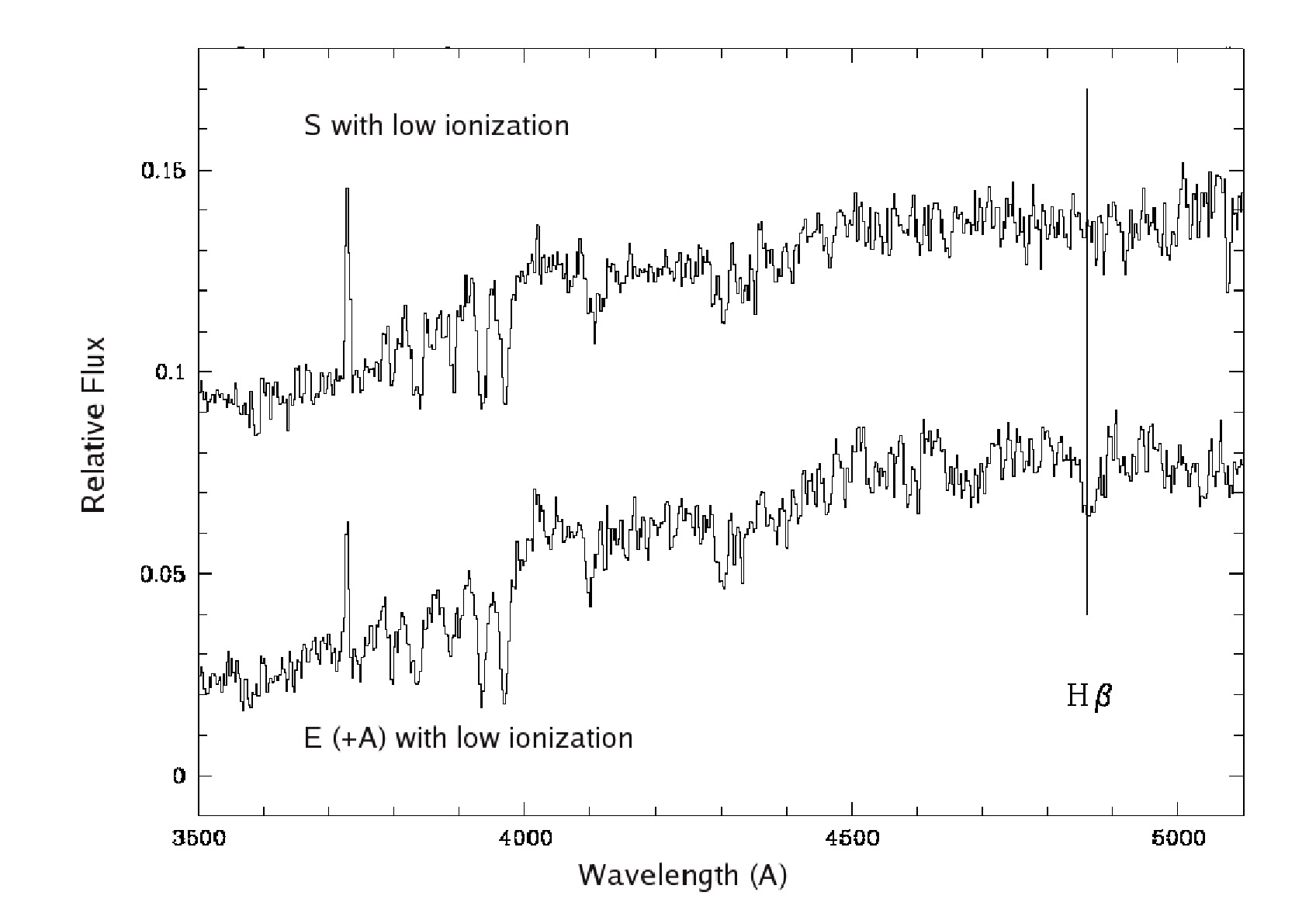}
    \vspace*{-.0cm}
    \caption{Median spectra of 11 early-type galaxies and of 10 galaxies with later type morphology 
(S) with low ionization at $z = 0.4 - 0.5$.
}
    \label{LINERS}
\end{figure}

Our visual early-type morphologies are the same as ZEST type T=1 \citep[][ Figure 4 (b), (c), (d)]{Scarlata:2007}.  
In the $<z> = 0.43$ bin we find that 92\% of the galaxies classified as star-forming objects have morphologies inconsistent with 
early-types. At $<z> = 0.43$ and in the $z = 0.63$ layer, about half of the LINERs have compact morphology while the other
half are mainly bulge-dominated disk galaxies, or ``early disks'' of ZEST type T=2.0 (\citep[][ Figure 4]{Bundy:2009}). 
At $z=0.29$ all LINERs have disks. The spectra of galaxies with apparent disks have an extended [OII] emission suggesting that they do 
have extended star-formation. Average spectra of 11 early-type galaxies (E) and 10 later types (hereafter S) resulting from our 
morphological classification are shown in Figure~\ref{LINERS}.



The absence of $\rm H\beta$ in the S sample  suggests that  $\rm H\beta$ in emission resulting from star formation is diluted in $\rm H\beta$ in absorption from A and older stars. The closest spectral comparison in the atlas of galaxy spectra \citep{Kennicutt:1992p5569} is with an Sb galaxy. The rather strong  $\rm H\beta$
in absorption in early-types (E) combined with [OII] suggests either a low fraction of young stars or a mechanism of photo-ionization other 
than young stars  as in \citep[][ and references therein]{Filippenko:2003, Ho:2004}. In fact, the recent work by the SEAGAL collaboration \cite[][
and references therein] {CidFernandes:2010} has shown that the majority of galaxies with LINER spectra in the SDSS can be explained as  {\it retired galaxies}, that is, galaxies that have stopped forming stars but still contain appreciable amounts of gas that is being photoionized by intermediate-aged post-AGB stars. In fact, the SEAGAL models with no young stars, but with significant populations of 100Myr-1Gyr stars resemble remarkably well our average LINER spectrum shown in Figure~\ref{SPEC0415lineseyfsf}.

\begin{figure}
   \centering
   \vspace*{0.5cm}
   \includegraphics[width=7cm]{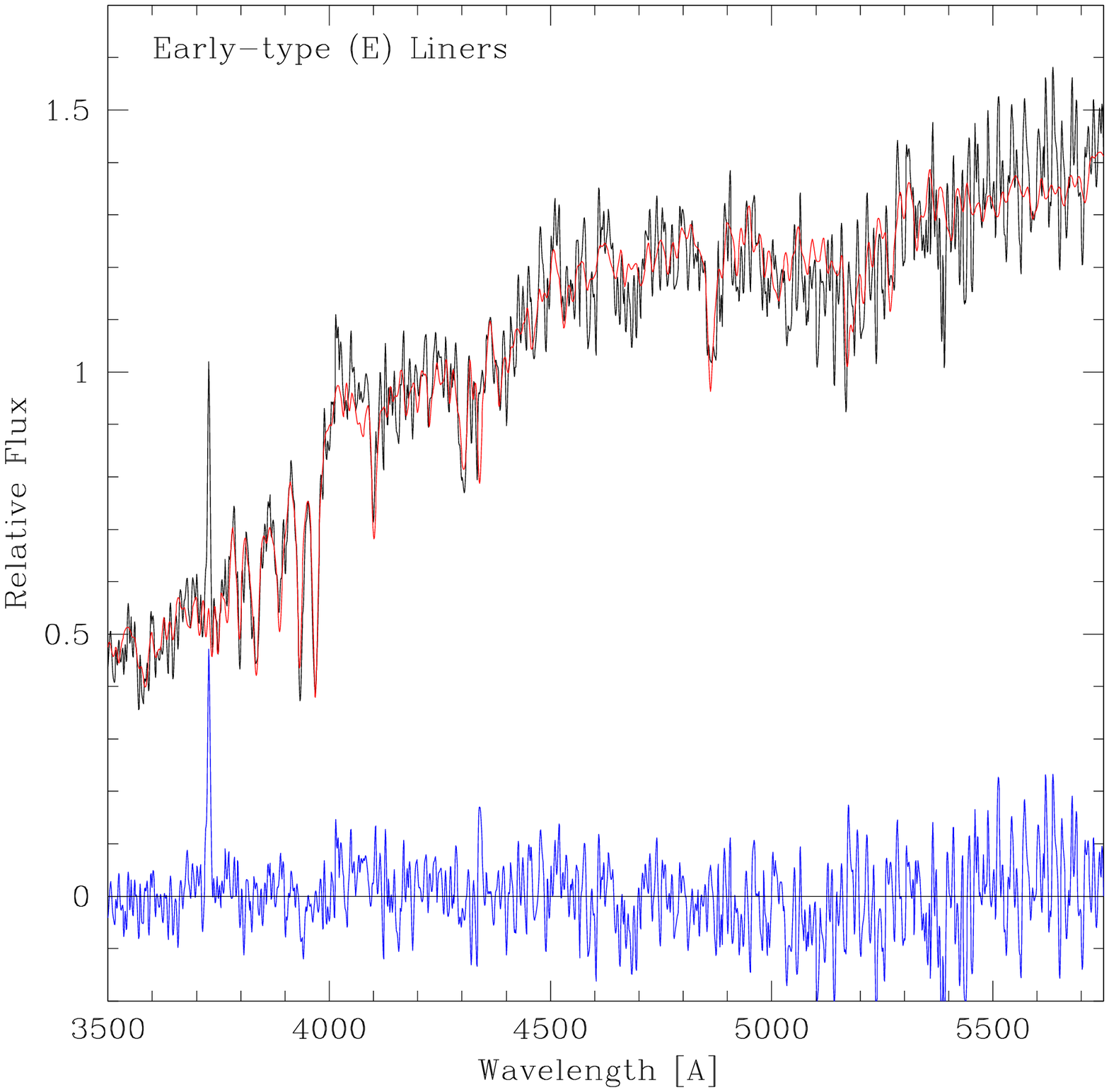}
   \includegraphics[width=7cm]{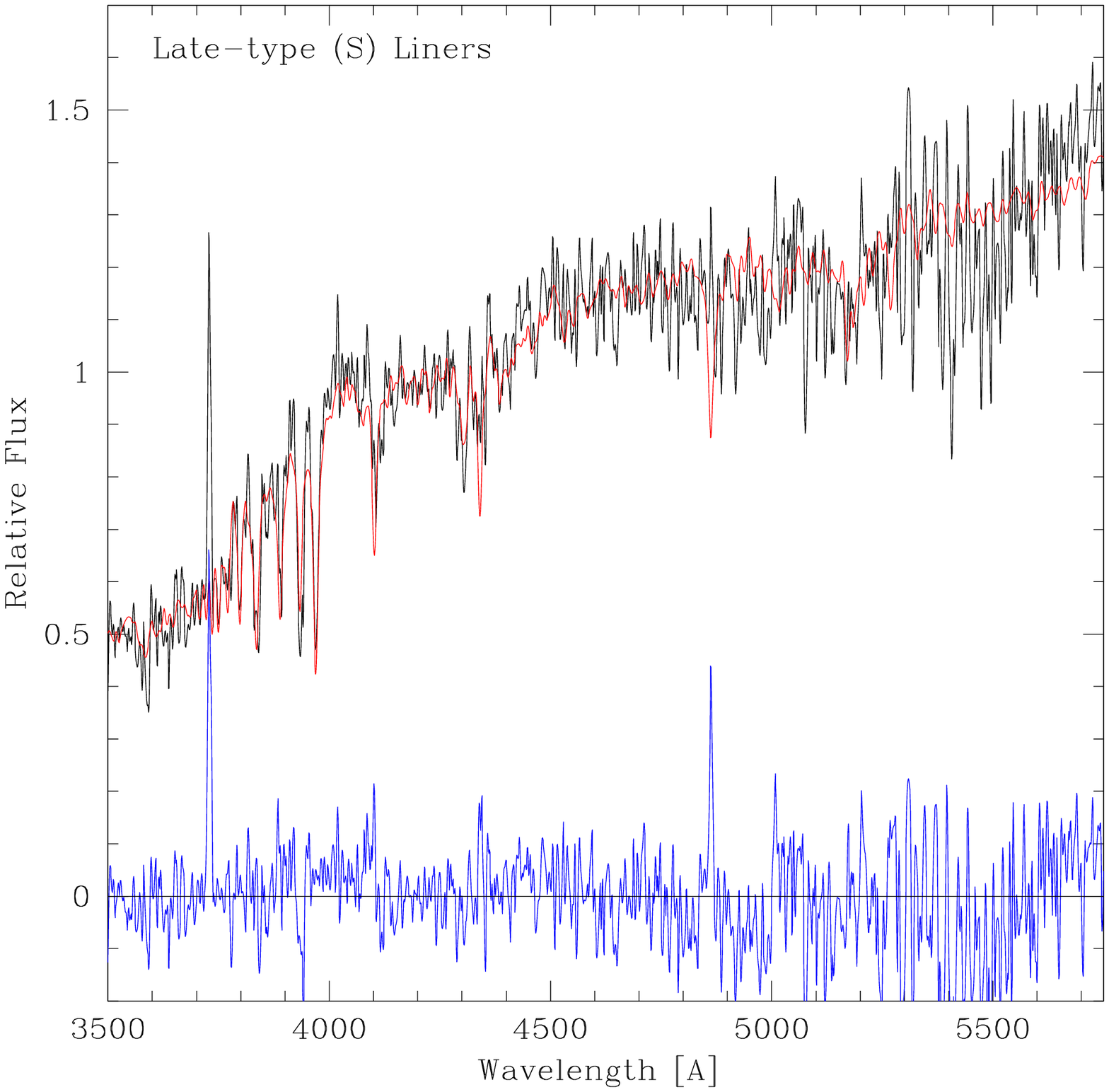}
    \vspace*{-.0cm}
    \caption{Spectral fitting with SSP models for an average spectrum
    of LINERs with early-type morphology (left), and with morphology of later type (right).}
    \label{liners_avg}
\end{figure}

We calculated population synthesis models for our average spectra of LINERs with early-type and late-type morphologies. The results, shown in Figure~\ref{liners_avg} and  Table~\ref{earlylinerpop}, indicate that  both early-type and late-type LINERs have significant populations of young and intermediate age stars, but late-type (S) LINERs have much younger populations. In fact, the residuals of the S-LINER fit show $\rm H\beta$ in emission stronger than  [OIII], consistent with the idea that they are red spirals with diluted star formation.


\begin{table*}
\renewcommand{\arraystretch}{0.9}
\centering 
\caption{Stellar population properties of an average of LINERs with early-type morphology and
with morphology of later types}
\begin{tabular}{ l l  c c c c c }
   Type  & \multicolumn{5}{c}{log Age} &$\rm \chi^2$ \\ 
           &   $<8$  & $8-8.7$ & $8.7-9$ & $9-9.4$ & $>9.4$   &  \\ \hline
           & & & & & & \\
              
    Early      &18.2    &  0   &        57.6   &       0     &      24.2   &  1.4   \\
    Later-type      &32.7   &  0   &        53.4   &       0     &      13.9   &   1.3    \\
\hline \hline
\label{earlylinerpop}
\end{tabular}
\end{table*}


Thus, our results are consistent with the interpretation that most early-type LINERs at intermediate redshifts are in fact post-starburst galaxies, as postulated  by the SEAGAL collaboration for lower redshift objects. These results indicate that LINERs and E+As depict the quenching phase in the evolution of galaxies massive enough to retain significant amounts of gas after the stellar-wind and supernova phases of the most massive stars. 
\rm

\subsection{The red limit of emission-line galaxies}
\label{redlimit}

At each $z$ we have selected galaxies with the reddest continuum (the reddest quartile at each redshift bin) to construct the combined spectra of the red envelope or red limit of emission line galaxies. Since we are working with small numbers of galaxies, typically 5-10, it was necessary to combine the samples at $z=0.82$ and $z=0.99$ to improve statistics.
Nevertheless, because our red emission-line galaxies are rather luminous, the combined spectra still have high continuum S/N ratios (Table~\ref{redest}). 
The common parts of the red envelopes of spectra at $<z>=0.29$, $<z>=0.43$, $<z>=0.65$ are similar, while the red limit at $<z>=0.9$ has noticeably stronger UV continuum. The spectra in different bins are shown in Figure 7 of Giraud et al. (2010). Both the continuum and the indexes of the red limit at $ z\leq0.65$ 
(i.e. $\rm D(4000)\sim1.35-1.45;  EQW([OII])\sim4-8$), are typical of nearby spirals with prominent bulges and low star formation \citep{Kennicutt:1992p5569, Kinney:1996p5582, Balogh:1999p4259}, or early-type LINERs. Up to $z\sim0.7$ the populations of red spirals and early-types can be well separated by their morphology.
The higher UV continuum and lower $\rm D(4000)$ of the limit  spectrum at $<z> = 0.9$ indicate that such red objects become rare at $z \geq 0.68$ in our sample. 

Absorption systems have (by definition) already lost enough gas to suppress any detected star formation by the time they first appear in our sample at $z\simeq1$.  At $z =0.8-1$ emission-line galaxies in our sample are found to have very strong star formation, which declines at lower $z$, the reddest quartile being bluer than at lower $z$. Therefore the evolutionary paths of bright absorption and emission systems might have been more separated at $z\simeq1$ than at lower redshift suggesting two different physical processes of different time scales. In one we have early-type LINERs and E+A galaxies that define the "entrance gate" to the red sequence of passively evolving galaxies. In the other we have red spirals with diluted star formation, that are in a final phase of smooth star formation, possibly of a ``main sequence'' \citep{Noeske:2007}.

In  Section~\ref{liners} we found a large fraction of LINERs in layers at intermediate  $z$. More precisely, in the volume-limited range $0.35\leq z\leq0.65$, we find, gathering the counts of Section~\ref{liners}, that LINERs are $\rm 23 \%$ of all early-type galaxies with measured redshifts. 


\begin{table*}
\renewcommand{\arraystretch}{0.9}
\centering 
\caption{Equivalent width of [OII], 4000~\AA\ break amplitude,  $\rm H\delta$ index, and
the G-step  of the red envelope of emission-line galaxies. The continuum S/N ratios are given in the last column. \rm}
\begin{tabular}{ c c c c c c }
$<z>$      &  EQW([OII])        &  D(4000)        & $\rm EQW(H\delta)$ & G step               & $\rm S/N$  \\ \hline
& & & & & \\
0.29       &  $4.2 \pm 0.3$     & $1.35 \pm 0.07$  & $-1.8 \pm 0.3$     &  $1.224 \pm 0.017$  & $\rm 19$      \\ 
$0.43$   &  $8.5 \pm 0.2$     & $1.39 \pm 0.06$  & $-2.5 \pm 0.2$     & $1.268 \pm 0.014$     & $\rm 24$ \\
$0.65$   &  $8.3 \pm 0.2$     & $1.44 \pm 0.06$  & $-3.0 \pm 0.2$     & $1.273 \pm 0.012$     & $\rm 29$ \\
$0.9$    &  $9.0 \pm 0.3$     & $1.30 \pm 0.07$  & $-3.9 \pm 0.2$     &  -                    & $\rm 18$ \\
\hline \hline
\label{redest}
\end{tabular}
\end{table*}


\section{Summary and Conclusions}

We have presented a catalogue of galaxy spectra in a pencil beam survey of $\sim 10.75' \times 7.5'$, and used
these data to make an analysis of the spectral energy distribution of a magnitude limited sample up to $z \sim 1$,
concentrating on absorption and low ionization emission-line systems.
The redshift range has been divided in bins centered on the structures that were detected in the (RA, Dec, $z$)  pseudo-volume, and corresponding to cosmic time slices  of $\rm \sim 1 Gyr$. Our sample is reasonably complete for galaxies brighter  than  $\rm M_R=-18.8$ up to $z \approx 0.5$;  at $z \geq 0.75$ the cutoff is at -20.5.

From this analysis we reach the following conclusions:

\begin{enumerate}

\item We confirm in our pencil-beam sample the well known result \cite{Hamilton:1985} that absorption-line galaxies do not show significant variations
in their continuum energy distributions up to $z=0.6$, and a moderate decrease of the 4000~\AA\  break amplitude of 5\% at $z\sim0.65$, 7\% at $z\sim0.82$, and up to 15\% at $z\sim1$. Using stellar population synthesis models we find that absorption-line galaxies at $z\geq0.65$ show more than 50\% of stars younger than 2.5Gyrs, while those at $z\geq0.8$ had star formation as recently as 1Gyr ago. This suggests that the red sequence is still in a buildup phase at $z\leq1$.

The faint absorption-line galaxies in our dynamically young cluster at  $z=0.29$ have indexes similar to those of bright absorption-line systems at $z=0.8$, suggesting that faint galaxies without emission lines tend to be younger than  more massive galaxies with similar spectra.  Our population synthesis models indicate that about 50\% of the stars contributing 
to the luminosity of faint absorption-line galaxies at z = 0.29 were formed at $z < 1$. This is consistent with cases of truncated red sequences observed in some 
high-$z$ clusters and suggests that clusters with truncated red-sequences may be dynamically young. 

\item Combining simple emission-line diagnostics with galaxy morphology we identify a significant population of early-type LINERs at $0.35\leq z\leq0.65$. In that redshift range  early-type LINERs constitute about 23\% of all early-types galaxies, a much larger fraction than E+A post-starburst galaxies. However, our population synthesis models show that early-type LINERs contain substantial populations of intermediate age stars that can easily explain the observed line emission, as recently proposed by Cid-Fernandes et al. (2010). This led us to conclude that most LINERs in  our sample are in fact post-starburst galaxies. 

\item The red limit in the spectral energy distribution of emission-line galaxies at $z\leq0.65$ is typical of bulge-dominated spirals with moderate 
star formation, and of early-type LINERs.  Thus,  early-type LINERs and E+As define the ``entrance gate''  of the red sequence of passively evolving galaxies,
while bulge-dominated spirals have diluted star formation.

\end{enumerate}

\begin{acknowledgements}
 EG thanks the hospitality of ESO and  Universidad Catolica in Santiago during the initial phase of this work.
 JME thanks the hospitality of Nanjing University during the initial phase of this research.
 QGU would like to acknowledge the financial support from the China Scholarship
 Council (CSC),  the National Natural Science Foundation of China under
 grants 10878010, 10221001, and 10633040, and  the National Basic Research Program 
 (973 program No. 2007CB815405).
 HQU thanks partial support from FONDAP ``Centro de Astrof\'isica''.
 PZE acknowledge a studentship from CONICYT. We thank S. di Serego Alighieri for reading a preliminary version
of the manuscript and for his suggestions, and R. Cid-Fernandes for fruitful discussions.
\end{acknowledgements}

\end{document}